\documentclass[12pt,dvipsnames]{iopart}

 \expandafter\let\csname equation*\endcsname\relax
 \expandafter\let\csname endequation*\endcsname\relax
\usepackage{amsmath,amsfonts,amssymb}

\usepackage{graphicx,subfloat}
\usepackage{color}
\usepackage{comment}
\usepackage{appendix}
\usepackage{multirow}
\usepackage{epstopdf}
\usepackage{etoolbox}
\usepackage{hyperref}

\makeatletter
\renewcommand\@appendixstar{\@@par
 \ifnumbysec
 \@addtoreset{table}{section}
 \@addtoreset{figure}{section}\fi
 \setcounter{section}{0}
 \setcounter{subsection}{0}
 \setcounter{subsubsection}{0}
 \setcounter{equation}{0}
 \setcounter{figure}{0}
 \setcounter{table}{0}
 \def\thesection{\Alph{section}} % this line has been \def\thesection{Appendix \Alph{section}} before
 \def\theequation{\ifnumbysec
      \Alph{section}.\arabic{equation}\else
      \Alph{section}\arabic{equation}\fi}
 \def\thetable{\ifnumbysec
      \Alph{section}\arabic{table}\else
      A\arabic{table}\fi}
 \def\thefigure{\ifnumbysec
      \Alph{section}\arabic{figure}\else
      A\arabic{figure}\fi}}
\makeatother
\begin{document}
\graphicspath{{figs/}}
\renewcommand{\d}{\text{d}}
\newcommand{\sieq}{\hspace{0.5cm}}
\newcommand{\changed}[1]{{\textcolor{black}{#1}}}

\title{Anomalous \changed{spectral laws} in differential models of turbulence}
\author{Simon Thalabard$^1$, Sergey Nazarenko$^{2,3}$,  S\'ebastien Galtier$^4$ and Sergey Medvedev$^5$
}

%\affiliation
\address{$^1$  Department of Mathematics and Statistics, University of Massachusetts, Amherst, MA 01003, USA. \\
 %Laboratoire Lagrange UMR 7293, Universit\'e de Nice-Sophia Antipolis,
%CNRS, Observatoire de la C\^{o}te d'Azur, Bd. de l'Observatoire, 06300 Nice, France. \\
$^2$ Mathematics Institute, University of Warwick, Coventry CV4 7AL, United Kingdom \\
$^3$ Service de Physique de l'Etat Condens\'e (DSM/IRAMIS/SPEC) - CNRS/MPPU/URA 2464,
B\^at 772, Orme des Merisiers,
CEA Saclay,
F-91191 Gif sur Yvette Cedex,
France \\
$^4$ LPP, \'Ecole Polytechnique,
91128  Palaiseau Cedex,
France \\
{$^5$ Institute of Computational Technologies SD RAS,
Lavrentjev avenue 6, Novosibirsk 630090, Russia}
}
\date{\today}

\begin{abstract}
Differential models for  hydrodynamic, passive-scalar and wave turbulence given by  nonlinear first- and  second-order  evolution equations  for the energy spectrum in the $k$-space were analysed.
Both types of models predict formation an anomalous  transient power-law spectra.
The second-order models were analysed in terms of self-similar solutions of the second kind, and  a phenomenological  formula for the anomalous spectrum exponent was constructed using numerics for a broad range of parameters covering all  known physical examples.
The first-order models were examined analytically, including finding an analytical prediction for the anomalous exponent of the transient spectrum and description of formation of the Kolmogorov-type spectrum as a reflection wave from the dissipative scale back into the inertial range. The latter behaviour was linked to  pre-shock/shock singularities similar to the ones arising in the Burgers equation.
Existence of the transient anomalous scaling and the reflection-wave scenario are argued to be a robust feature common to the finite-capacity turbulence systems. The anomalous exponent is independent of the initial conditions but varies for for different models of the same physical system.
\end{abstract}

\maketitle
\section{Introduction}
Differential models of turbulence, such as the Leith  model (\cite{Leith67}) and
 the  Kovasznay  model  (\cite{Kovasznay48}) have played an important role for achieving a qualitative and even quantitative understanding of turbulence in various physical situations.
Differential models have been also used in wave turbulence. In particular, for the water gravity wave turbulence such a model was introduced in
\cite{hasselman}.
The main advantages of the differential models are their relative simplicity and their great flexibility when it comes to the  inclusion of new physical effects, \emph{e.g.} simultaneous presence of a strong and a weak turbulence components in superfluid turbulence (\cite{gradual}), modelling  atmospheric turbulence with two scale-separated sources
(\cite{lilly1989two}), new types of forcing or dissipation like  reconnections of superfluid vortices (\cite{nazarenko2007kelvin})
(\cite{nazarenko2006KWDAM}) or sound radiation by vortices (\cite{nazarenko2006KWDAM}).

Another advantage of the differential models is that they allow to study non-stationary turbulence.
Let us consider an isotropic hydrodynamic turbulence with an initial spectrum in a finite range of wave numbers $k = |{\bf k}|$. How will such a system evolve toward the Kolmogorov power-law spectrum with a  $k$-independent flux of energy in the inertial range of scales?
One could guess that the spectrum would spread out of the initial finite range of scales toward high values of $k$ leaving the Kolmogorov spectrum immediately behind the moving front. However, recent findings obtained within the integro-differential  and differential closures of turbulence, as well as the exact results for the Burgers equation, indicate that this may be an incorrect view. By solving numerically the kinetic equation for weak MHD turbulence, it was found in \cite{galtier2000weak} that the initial spectrum behind the moving front is indeed a power law, $E \sim k^{-x}$, but it has an exponent $\approx 7/3$ which is clearly distinguishable and steeper than the one of the Kolmogorov-type spectrum ($2$ for the weak MHD turbulence). It was found that the Kolmogorov-type spectrum gets formed in the system by a reflection wave propagating from high to low wave numbers after the spectral front reaches the dissipation scale. Similar behaviour was observed for the differential (Leith) and integro-differential (EDQNM) closures of Navier-Stokes turbulence in \cite{Connaughton2004} and \cite{bos2012developing} respectively.
Rigorous bounds for the anomalous index in the Leith model were recently found in \cite{Grebenev2014}: the index was proven to be strictly greater that Kolmogorov's $5/3$ and strictly less than $1.95$ which agrees with the numerical value $\approx 1.85$.

For the Burgers equation, it was noted long ago in \cite{sulem1983tracing} that the typical wave breaking scenario in which a pre-shock (cubic root singularity) is followed by a shock (jump discontinuity) in the $k$-space corresponds to an initial formation of a  power-law spectrum with exponent $8/3$ which is gradually replaced by the Burgers $x=2$ spectrum invading the $k$-space as a wave moving from high to low $k$.

Presently, it is understood that the described above scenario in which an anomalous power-law precursor spectrum appears prior to the Kolmogorov-type spectrum must be a generic property of the finite capacity systems, i.e. the systems for which   the Kolmogorov-type spectrum is integrable at $k =\infty$ (meaning that for vanishing dissipation  the Kolmogorov-type spectrum contains only a finite energy).
This scenario is described, using the  classification of \cite{BARENBLA_ZELDOVIC}, by a self-similar solution of the second kind.
The fact that a turbulent system belongs to the finite capacity type does not depend on the turbulent closure and, therefore, we should expect that the exact system exhibits the same
anomalous scaling behaviour without any assumptions or approximations related to a particular closure. However, the levels of resolution and precision of the current experimental measurements and numerical simulations are currently insufficient to  identify a  scaling range prior to formation of the Kolmogorov spectrum
which would be wide enough to be able to distinguish confidently between the Kolmogorov and the anomalous exponents.
On the other hand, the anomalous power-law index, being independent of the initial conditions, does depend on the closure used, as will be seen in the MHD and Navier-Stokes turbulence examples later in this paper.  Whether there is a universal value for
the anomalous index and if such a value could be traced to particular singularities of the dynamical fluid equations (like in the Burgers equation example) remain questions to be answered.

In the present paper we extend the study of the anomalous power-law to a general three-parametric class of the differential models, which is relevant to several known examples of strong and weak finite capacity turbulence systems.
In addition to the second-order equations of the nonlinear diffusion type generalising the  model of \cite{Leith67}, we study simpler first-order nonlinear wave equations generalising the  turbulence model of \cite{Kovasznay48}. Those  second-order models can be solved exactly. They yield an  anomalous power-law scenario, that may  be explained by the formation of Burgers-type singularities -- pre-shocks and shocks.

The paper is organised as follows. In section \ref{sec:Nonlinear diffusion models for turbulence}, we describe a generalized version of Leith Model that can be used as a toy model for turbulent systems with a finite-capacity spectrum. In section \ref{sec:Self-similar solutions of the nonlinear diffusion models}, the calculations of \cite{Connaughton2004} and \cite{Grebenev2014} are extended to infer the  existence of self-similar solutions whose energy spectra are steeper than Kolmogorov's. In section \ref{sec:Numerics}, we use numerics to characterize this anomaly, and construct a phenomenological fitting formula for the anomalous exponent as a function of the parameters of the model.  In section \ref{sec:The first-order model}, we discuss  an even simpler class of models which include the  Kovasznay model of 3D turbulence \cite{Kovasznay48}, and for which anomalous exponents exist and can be analytically derived. Finally, in section \ref{sec:summary} we present a summary and discussion of our results.

%\section{Differential models of turbulence}
\section{Nonlinear diffusion models for turbulence}
\label{sec:Nonlinear diffusion models for turbulence}

We consider the following inhomogeneous non-linear diffusion procfirst-orderess in $k$-space for the energy spectrum $E(k,t)$,
\begin{equation}
 \partial_t E = C\, \partial_k \left[k^mE^n\partial_k(E/k^{d-1})\right],
\label{eq:leithmodeldim}
\end{equation}
where $d =1, 2$ or $3$ is the dimension of the system and indices $m$ and $n$ are real (usually rational positive) numbers depending on the physical system. Constant $C$ in general has a physical dimension which depends on the particular system.

The model (\ref{eq:leithmodeldim}) is constructed so that it has two fundamental  power laws $E \sim k^{-x}$ as stationary solutions -- a constant flux Kolmogorov-like spectrum with
$x=x_K = (m-d)/(n+1)$
and a thermodynamic equilibrium spectrum with {$x=x_T=1-d$}.

One can check that the diffusion model also admits the general time-independent solution
\begin{equation}
\label{eq:warmCas}
%E_s(k) = k^2 \left(Ak^{1-2n-m}+B \right)^{1/(1+n)}
E_s(k) = k^{d-1} \left(Ak^{1-m + (1-d)n}+B \right)^{1/(1+n)}
\end{equation}
with arbitrary positive real constants $A$ and $B$. {Case $B=0$} corresponds to the Kolmogorov-type spectrum and {case $A=0$} to the thermodynamic spectrum. Thus,
the case with both $A$ and $B$ having non-zero values is a mixed state which was named a ``warm cascade" in \cite{Connaughton2004}.
Note that  at  $k \to 0$, the mixed state (\ref{eq:warmCas}) is dominated by the Kolmogorov (flux) part and for $k \to \infty$, this solution tends  the thermodynamic equilibrium spectrum.

Note that equation  (\ref{eq:leithmodeldim}) describes a conservative model, $\int_0^\infty E \, dk =$~const. Naturally, viscosity can be taken into account by adding to the right-hand side of this equation term $-\nu k^2 E$, where $\nu=$~const~$>0$ is a viscosity coefficient. Other types of dissipations may also be modelled depending on the physical system. However, for the second-order models we will only be concerned with non-stationary initial states of evolution at which the dissipative scales are not reached and, therefore, the dissipative terms can be neglected. (We will be able to have a more general treatment including the stage where the viscosity is important when we consider the first-order models).

Also, in this paper we will be concerned with {\em finite capacity} systems for which the Kolmogorov-type spectrum is integrable at $k \to \infty$. In the other words, no matter how large the dissipation wave number is, the Kolmogorov state will have a finite energy density in the physical space. Thus, we are interested in the physical examples with $x_K = (m-d)/(n+1) > 1$. Let us now mention some examples.

\subsection{Examples}
\label{ssec:leithexamples}
\subsubsection{3D hydrodynamic turbulence}

The  case with $d=3, m=11/2$ and $n=1/2$ corresponds to the Leith model  introduced in \cite{Leith67}. In this case constant $C$ is dimensionless and has value $\approx 1/8$ (\cite{Connaughton2004}).

\subsubsection{Passive scalar turbulence}

In this case $E$ has the meaning of the passive scalar spectrum and $d=2$ or $3$. Also, since the advection is passive, the resulting equation must be linear in $E$, therefore $n=0$.
The constant $m$ depends on the roughness of the advecting velocity field.
For smooth velocity (Batchelor's regime) we have $m=d+1$ so that $x_K = 1$ (\cite{Batchelor59}). Actually, the Batchelor regime is on the borderline of the infinite capacity systems and strictly speaking our analysis will be inapplicable to this case. However, we will see that the case $x_K = 1+\epsilon$ is interesting and it will nicely illustrate a smooth transition between the finite and the infinite capacity regime asymptotics.

For a rough velocity corresponding to the Kolmogorov spectrum (Obukhov-Corsin regime) we have $m=d+5/3$ so that $x_K = 5/3$ (\cite{Obukhov49,Corrsin51}).

\subsubsection{Wave turbulence }

In this case  $d=1,2$ or $3$ and $n$ is the order of the resonant wave interaction less two, e.g. $n=1$ for three-wave processes, $n=2$ for four-wave processes, etc.
Indeed, the nonlinearity degree of the differential equation is $n+1$ i.e. the same as the one of the respective integro-differential wave-kinetic equation.
The constant $m$ depends on the particular type of the waves. Particular examples include:

\begin{enumerate}
\item
{\em Gravity water waves:} In this case $d=2$, $n=2$ (a four-wave processes) and $m=19/2$ so that $x_K = 5/2$ (see \cite{Zakharov1967}; for the differential model for this system see \cite{ZakharovPushkarev1999} and \cite{nazarenko2006sandpile}).
\item
{\em Capillary water waves:} In this case $d=2$, $n=1$ (a three-wave processes) and $m=11/2$
so that $x_K = 7/4$  (see e.g. \cite{Pushkarev200098}).
\item
{\em Sound  waves:} In this case $d=3$, $n=1$ (a three-wave processes) and $m=6$ so that $x_K = 3/2$ (see \cite{1970zs}; the differential approximation for this system is the same as the one for isotropic MHD turbulence derived in \cite{Iroshnikov}).
\item
{\em  Alfven waves:} In this case $d=2$ (because even in 3D the system is very anisotropic and the spectrum is almost two-dimensional) $n=1$ (a three-wave processes) and $m=6$ so that $x_K = 2$ (see \cite{galtier2000weak}; for the differential model for this system see \cite{galtier2010nonlinear}).
\item
{\em Kelvin waves on vortex filaments:} In this case $d=1$, $n=2$ (a four-wave processes) and $m=6$ so that $x_K = 5/3$ (see \cite{l2010weak} and \cite{boue2011exact}).

\end{enumerate}

\begin{comment}
\begin{equation}
 \partial_t E = \partial_k \underbrace{\left[k^mE^n\partial_k(E/k^2)\right]}_{\Pi(k)}.
\label{eq:leithmodel}
\end{equation}
\end{comment}
%

\section{Self-similar solutions of the nonlinear diffusion models}
\label{sec:Self-similar solutions of the nonlinear diffusion models}

We are interested in the early evolution of the spectrum which is initially non-zero only in a finite range of $k$. As the front of the spectrum propagates toward large $k$, for the values of $k$ which are much greater than the initial $k$'s the solution tends to
a self-similar solution of the second kind. It is to be found by making in  equation  (\ref{eq:leithmodeldim}) a substitution in the form
\begin{equation}
 E(k,t)= (t_\star-t)^\alpha F(\eta),\sieq  \eta=k/k_\star, \sieq
\text{and} \sieq k_\star=c(t_\star-t)^\beta.
\end{equation}
Such a solution describes an explosive propagation of the spectral front $k=k_*$ in a finite time $t_\star$, i.e. $k_\star \to \infty $ as $t_\star \to \infty $ (function $F(\eta)$ is zero at $\eta>1$). Thus, the solution exists only for a finite time after which one can no longer ignore viscosity to describe the subsequent evolution.

Obviously the constant $c$ depends on the initial spectrum: the stronger turbulence is the faster it  evolves.
Solution for  the case with $c \ne 1$ can be obtained from the solution for the case with $c=1$ by a rescaling. Thus, thereafter we will put $c=1$. We will also put $C=1$ in
equation  (\ref{eq:leithmodeldim}) because it can be absorbed into the time variable.
To make the self-similar framework consistent, or in other words to make  the equation for $F$ involve the similarity variable $\eta$ only,  $\alpha$ and $\beta$ need to be chosen so that
%\begin{equation}
%n\alpha+ (m-4)\beta+1 =0
%\label{eq:abcondition}
%\end{equation}
%
\begin{equation}
\label{alphabeta}
n\alpha + (m-d-1) \beta +1=0.
\end{equation}
For a steady state power-law spectrum, $E \sim k^{-x}$ the self-similarity implies $F \sim  \eta^{-x}$. The meaning of such power-law asymptotics will become clear shortly. For now, we express $\alpha$ and $\beta$  in terms of  $x = - \alpha/\beta$ using the condition
(\ref{alphabeta}):
\begin{equation}
\alpha = - \frac x { nx+d+1-m}\sieq \text{and} \sieq \beta =  \frac 1 { nx+d+1-m}.
\end{equation}
We then obtain the following equation for the profile function $F$:
\begin{equation}
\dfrac{1}{d+1-m+nx} \left[xF+\eta\partial_\eta F \right] = \partial_\eta\left[\eta^mF^n\partial_\eta (F/\eta^{d-1})\right].
\label{eq:profiledim}
\end{equation}
The special case corresponding to the Leith model,  $d=3, m=11/2$ and $n=1/2$, was treated in \cite{Connaughton2004} and in
\cite{Grebenev2014}. In this case, $1/(d+1-m+nx) = 2/(x-3)$.
Below, we will extend the analysis to the general finite capacity case.

Following  \cite{Connaughton2004} and
\cite{Grebenev2014}, we realise that any initial spectrum concentrated in a finite range of
$k$ will asymptotically tend to a  self-similar solution for large $k$. This corresponds to
a solution of Equation (\ref{eq:profiledim})  with  two boundary conditions:
$F(1) =0$ corresponding to a sharp propagating front of the spectrum and
$F \sim \eta^{-x} $ at $\eta \ll 1$ corresponding to a stationary power-law spectrum forming behind the propagating front. This formulation is a nonlinear eigenvalue problem because it has a solution only for one value of $x=x_\star$. It is precisely the ``eigenvalue"  $x_\star$ and its dependence on $m,n$ and $d$ which we will aim to find below.

Like in  \cite{Connaughton2004} and
\cite{Grebenev2014}, we will first find an autonomous system equivalent to   equation  (\ref{eq:profiledim}). \changed{In addition to being autonomous, the new system must have  non-singular fixed points.  This is achieved by choosing parametrically  a suitable new time variable $\tau(F, \eta)$
.} The latter can be found first using the following change of variables,
%\begin{equation}
% F=\sigma^\mu \eta^{-\lambda} \sieq \text{and} \sieq F^\prime(\eta) = \lambda\eta^{-\lambda-1}\sigma^{\mu-1}\rho ,
%\label{eq:changevariable}
%\end{equation}
{
\begin{equation}
 F=\rho^\mu \eta^{-\lambda} \sieq \text{and} \sieq F^\prime(\eta) = \lambda\eta^{-\lambda-1}\rho^{\mu-1}\sigma,
\label{eq:changevariable}
\end{equation}}
where $\mu$ and $\lambda$ must be selected so that one gets a suitable autonomous
system {(see Appendix \ref{ap:autonomous}), namely :
\begin{equation}
 \mu =1/n  \sieq \text{and} \sieq  \lambda=(m-d-1)/n.
 \label{eq:lambdamu}
\end{equation}
It is now convenient  to work with the new independent variable $\tau = \tau(\eta)$, which we define  according to the formula
\begin{equation}
 \dfrac{ \d \tau}{\d \log \eta } = \dfrac{1}{\rho }, \sieq \text{so that} \sieq  \dfrac{ \d }{\d \eta} = \dfrac{1}{\rho \eta} \, \dfrac{\d}{\d \tau}.
 \label{eq:taueta}
 \end{equation}
Plugging these changes of variables  into Equation  (\ref{eq:profiledim}) then  yields  the following autonomous system :}
\begin{eqnarray}
 \rho^\prime(\tau) &=& \dfrac{\lambda}{\mu}\rho(\rho+\sigma), \label{eq:evodim1}
\\
 \sigma^\prime(\tau)  &=&  \dfrac{\mu(x\rho+\lambda \sigma)}{\lambda(x-\lambda)}-\lambda\sigma^2 + (d-1)\left(\dfrac{1}{\lambda} + \dfrac{1}{\mu} \right) \rho^2 + \left(\dfrac{d-1}{\mu} + d-2 \right) \rho \sigma.
\label{eq:evodim2}
\end{eqnarray}
There exist three fixed points for the system (\ref{eq:evodim1}) --  (\ref{eq:evodim2})  :
\begin{equation}
P_1=(0,0), \sieq P_2 = \left(0,\dfrac{\mu}{\lambda(x-\lambda)}\right), \sieq\text{and} \sieq P_3=(1,-1)  \dfrac{\mu}{(\lambda-1)(\lambda+d-1)}.
\label{eq:steadydim}
\end{equation}
The nature of the fixed points can be established by examining the Jacobian matrix $\Delta(\rho,\sigma)$ of system (\ref{eq:evodim1}) --  (\ref{eq:evodim2}), see appendix
\ref{ap:stability}.
We find that
$P_1$ is a saddle-node and $P_2$ is a saddle.
The nature of $P_3$ depends on the parameter values.
Namely, $P_3$ is an unstable node for $x\le x_-$, an unstable focus for $x \in[x_-;x_c]$ a stable focus for $x \in [x_c;x_+]$ and a stable node for $x \ge x_+$, where $x_\pm$ and
$x_c$ are explicitly defined in formulae (\ref{eq:xpm}) and (\ref{eq:xc}).
A local Hopf bifurcation of creation of a limiting cycle around $P_3$  occurs at $x=x_c$.
As found in  \cite{Connaughton2004} and
\cite{Grebenev2014}, the vicinity of the point $P_1$ corresponds to the $\eta \to 0$ part of the solution whereas point  $P_2$ corresponds to the sharp front, $\eta=1$.
The goal is to find such $x=x_\star$ that one could have an orbit connecting $P_1$ and $P_2$, i.e. a heteroclinic orbit.
Such a heteroclinic orbit arises at a global heteroclinic bifurcation at $x=x_\star$  associated with creation of a  heteroclinic cycle consisting of two heteroclinic orbits one of which being the required solution and the other one is the piece of the $\sigma$-axis connecting
$P_1$  and $P_2$ (see details in \cite{Grebenev2014}).

In the special case of the Leith model, \cite{Grebenev2014} showed that the value of $x_\star$ lies above $x_K$  and below $x_c$.  This result still holds true for the general class of models studied in the present paper. While this bounding can be used to obtain specific asymptotic behaviours for the critical exponent $x_\star$ (see Appendix \ref{ssec:asymp}), it does not indicate how  this exponent precisely varies for ``intermediate regimes'' of the parameters $(m,n,d)$.
Still, this theoretical bounding of $x_\star$ can be used as a first proxy to numerically determine $x_\star$ with a dichotomic search algorithm.
Our aim is twofold : \emph{i)} observe the general dependence of $x_\star$ with the parameters used in the model and \emph{ii)} provide a practical  ``engineering'' approximation for $x_\star(m,n)$ for the physically motivated values of the parameters $(m,n,d)$  previously described.
The algorithm that we use and the numerical results are described in the next section.

\section{Numerics}
\label{sec:Numerics}

\subsection{Dichotomic algorithm}
The value $x_\star$, such that the trajectory starts at $P_1$ and ends at $P_2$ can easily be determined numerically using a dichotomic algorithm. The basic idea is simple: if we start near $P_2$ and compute the trajectory backwards in $\tau$ for $x \ne x_\star$, we will miss $P_1$ and will end up either spiralling into the focus $P_3$ or crossing the $\sigma$-axis to the quadrant of positive $\rho$'s. Detecting these events will allow us to iterate $x$ closer to $x_\star$.

In the sequel, the set of equations  (\ref{eq:evodim1}) --  (\ref{eq:evodim2})  is rescaled so that $P_3=(1,-1)$ in  the
$(\rho,\sigma)$-plane.

\begin{enumerate}
 \item Set $x_1=x_-$ and $x_2=x_+$.
 \item $x \leftarrow (x_1+x_2)/2$.
 \item Compute $P_1$, $P_2$, $P_3$. Choose an initial point $P_i$ in the vicinity of $P_2$, say $P_i =P_2 + \dfrac{\epsilon}{\sqrt{2}}(1,1)$.
 \item Integrate equations  (\ref{eq:evodim1}) --  (\ref{eq:evodim2})  backwards in time until either  $\sigma(-\tau)>0$ or $|P(-\tau) - P_3 | \le \epsilon$ ($\epsilon \ll 1$ is prescribed). This sets a final time $\tau_f$.
 \item If $\sigma(-\tau_f) >0$ , then $x_2 \leftarrow x$ else $x_1 \leftarrow x$.
 \item Repeat steps (b) to (e) until $|x_2-x_1|<\eta$ ($\eta$  is a prescribed accuracy of the result).
 \item Return $x_\star = (x_1+x_2)/2$.
\end{enumerate}

\begin{figure}
\centering
\includegraphics[width=0.7\textwidth]{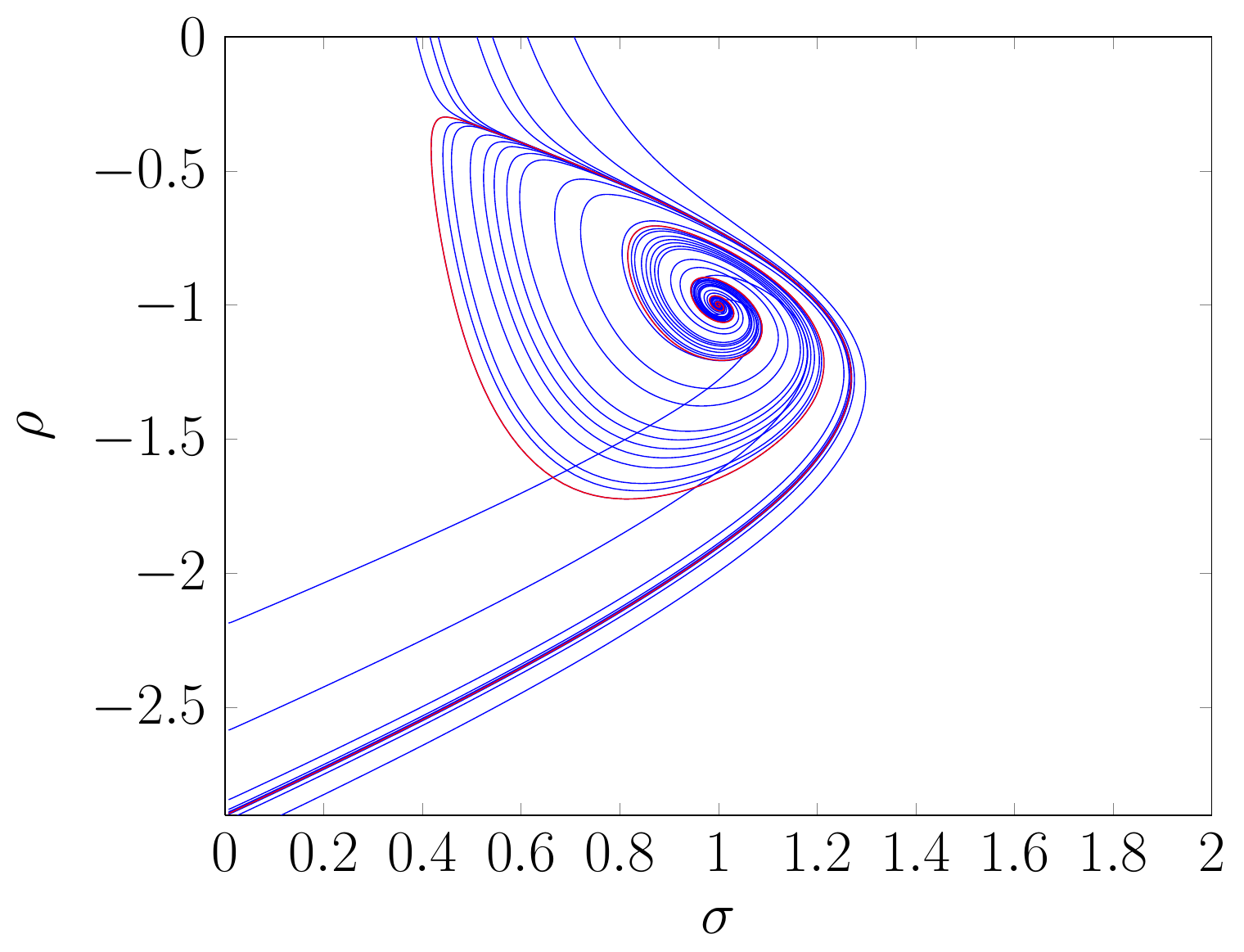}
\caption{Illustration of the dichotomic algorithm for $m=11/2$, $n=3/2$, $d=3$. Here $\epsilon=10^{-2}$ and $\eta=10^{-6}$ }
\label{fig:dicho}
\end{figure}

For $m=11/2$ and $n=3/2$, the algorithm yields $x_\star \simeq 1.85$; see Figure \ref{fig:dicho}.
\subsection{General Behaviour.}
Recall that the value of $(x_K,n,d)$ determines the triplet $(m,n,d)$ as $m=d-(n+1) x_K$. The dichotomic algorithm was run, in order to observe how the value of $x_\star$ varies with $x_K$, $n$ and $d$.
For each set of parameters, the value of $x_\star$ was determined with a precision of $\eta=10^{-5}$.\\
We considered $50$ different values of $n$'s logarithmically spaced between $e^{-3}\simeq 0.05 $ and $3$, and $50$ values of $x_K-1$'s logarithmically spaced between $e^{-5}$ and  $6$.
With this choice of parameters, we aim to investigate the behaviours $n \to 0^+$ which describes the passive scalar and $x_K \to 1^+$, the limit below which the energy spectrum has no longer a finite capacity. Note that we found that the dimension $d$ had no significative quantitative influence on $x_\star$, and therefore only discuss in details the numerical results obtained for the case $d=3$.

\begin{figure}
 \centering
\includegraphics[width=0.69\columnwidth]{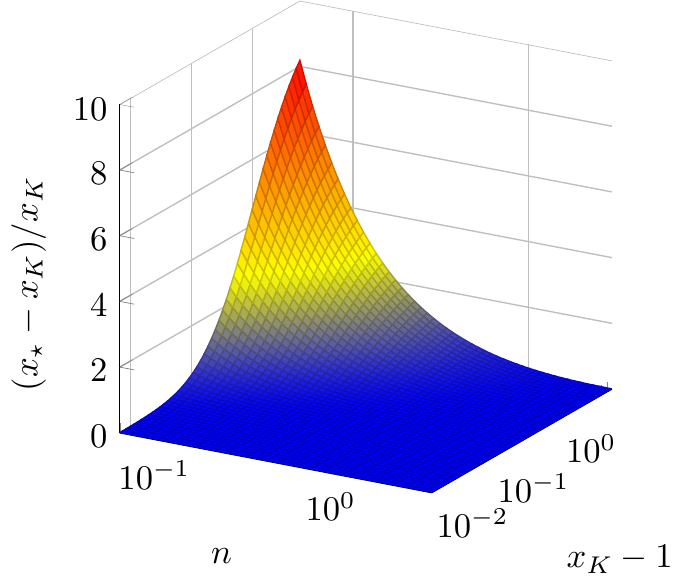}
 \caption{3D rendering of the deviation from the Kolmogorov exponent, measured through the quantity $\delta = (x_\star-x_K)/x_K$ plotted against $x_K$ and $n$ for the case $d=3$. The same plots for the cases  $d=1$ and $d=2$ would lie slighlty below, but would be practically undistinguishable from the case $d=3$.}
 \label{fig:n-x-dx}
\end{figure}

A 3D rendering of the general behavior of the deviation of the critical exponent $x_\star$ from  the Kolmogorov value $x_K$ is  displayed in Figure \ref{fig:n-x-dx}. The deviation is  here measured by the quantity $\delta x = (x_\star-x_K)/x_K >0$ and plotted against $x_K$ and $n$.  The figure indicates the following trends.
\begin{enumerate}
\item For fixed $n$, $|x_K -x_\star| \to 0$ as $x_K \to 1$.
\item For fixed $n $, $|x_K -x_\star| \to \infty$ as $x_K \to \infty$.
\item For fixed $x_K$, $|x_K -x_\star| \to 0$ as $n \to \infty$.
\item For fixed $x_K$, $|x_K -x_\star| \to \infty$ as $n \to 0$.
\end{enumerate}
Note that the asymptotic behaviors (a) and (c) observed in the numerics can be proven mathematically using the bounding theorem of \cite{Grebenev2014}. It suffices to recall that $x_-<x_K<x_\star<x_+$, explicitly compute $x_+-x_-$, and observe that this difference vanishes in both asymptotics. On the contrary, the behaviors (b) and (d) cannot be obtained with the same argument, as in the latter case the difference  $x_+-x_-$ diverge -- \, see Appendix \ref{ssec:asymp}. \\
It is interesting that the limits $x_K \to 1$ and $n \to 0$ do not commute.
Recall that the case $n = 0$ corresponds to the passive scalar turbulence and, in particular,
the case  $(n = 0, x_K = 1)$ corresponds to the passive scalar turbulence in Batchelor regime (smooth velocity). For the latter regime it is actually known that $x_K = x_\star$ in the case when turbulence is forced at large scales, \emph{i.e.} the
Kolmogorov-type (Batchelor-Kraichnan) spectrum develops right behind the propagating front; see \cite{FLM:15837}. In absence of forcing the spectrum is non-universal. This  is not surprising as this case describes a  (border-line) infinite capacity system. However, it is rather striking that in the passive scalar turbulence with rough velocity field $x_\star \to \infty$. This can  be interpreted as a spectrum that decays faster than any power law.
In other words, the Kolmogorov regime (Corrsin-Obukhov) gets established as a backward  wave propagating from the dissipative wave numbers  to the smaller wave numbers. There is no precursory scaling in the larger wave numbers in this case, \emph{i.e}. in the opposite direction with respect to the infinite capacity systems' behavior.
\subsection{An engineering fit for the physical range of parameters.}
\begin{figure}
 \centering
\begin{minipage}{0.49\textwidth}
 \centering
 \includegraphics[width=0.9\columnwidth]{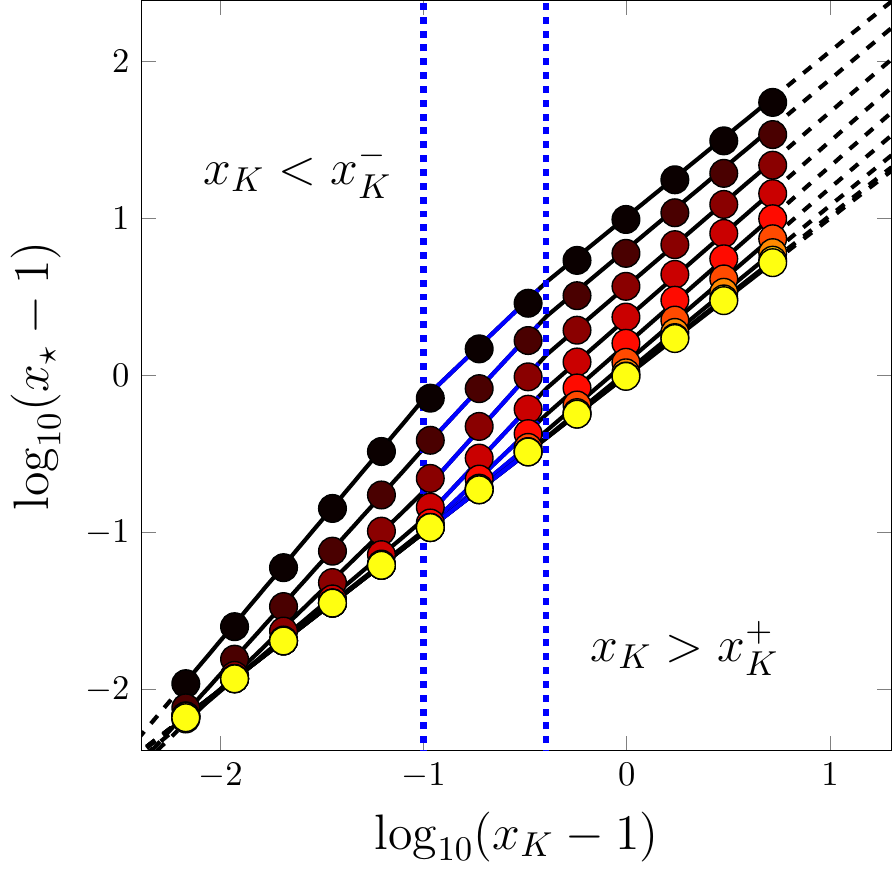}
\end{minipage}
\begin{minipage}{0.49\textwidth}
 \centering
 \includegraphics[width=0.9\columnwidth]{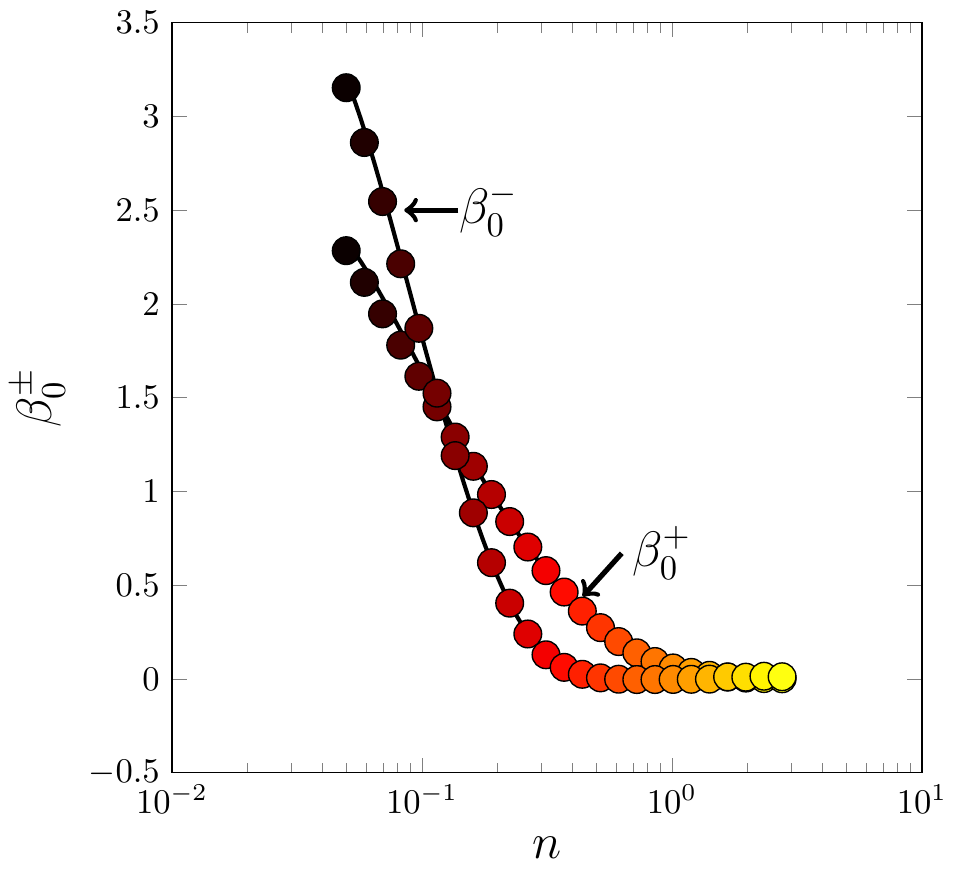}\\
 \includegraphics[width=0.9\columnwidth]{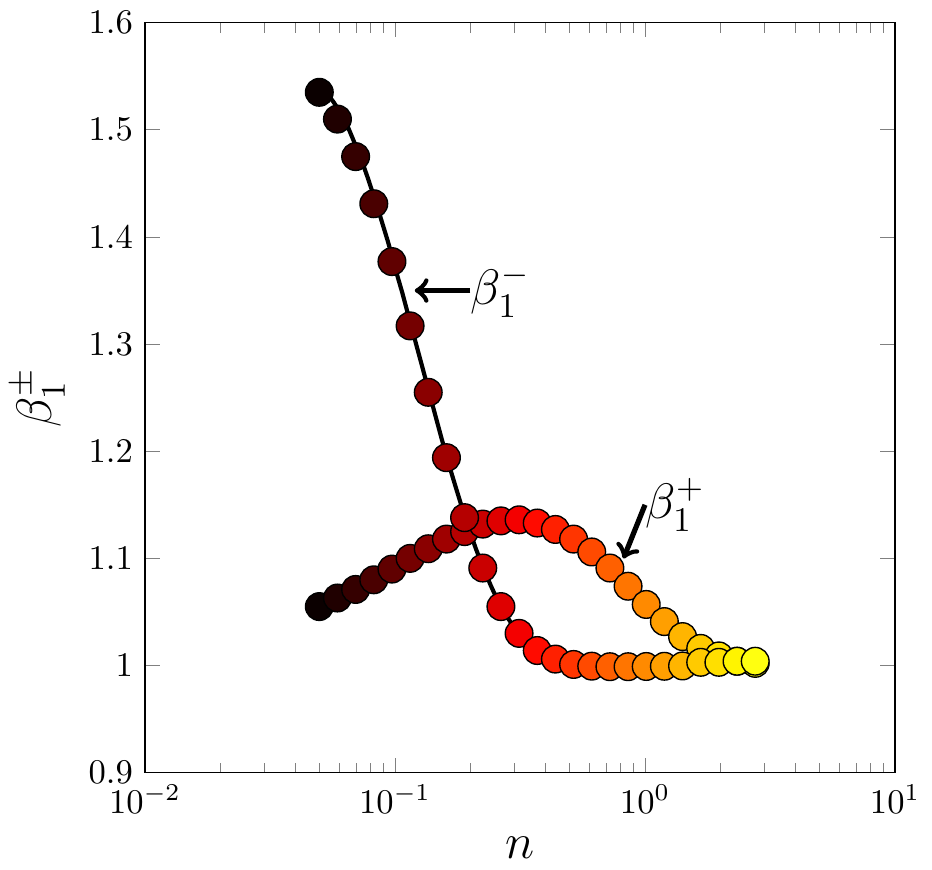}
\end{minipage}
 \caption{Left : Illustration of the piecewise-linear engineering formula (\ref{eq:engineeringfit}). The dots are the values obtained with the dichotomic algorithm, while the lines correspond to the formula (\ref{eq:engineeringfit}). For representation purpose, not all the data points are displayed. The color codes $n$ (see right panel). The blue dotted lines separate the regions $x_K < x_K^-$, $x_K^-<x_K<x_K^+$ and $x_K>x_K^+$. Indication of a $n$-dependent linear behavior is apparent for $x<x_K^-$ and $x >x_K^+$. Right : Evolution of $\beta_1^\pm(n)$ and $\beta_0^\pm(n)$ as measured from the dataset (see Formula (\ref{eq:engineeringfit}) for the definitions).   }
 \label{fig:engineeringfit}
\end{figure}
A closer analysis of our numerics  indicate the existence of $n$-dependent power-law behaviors for $x_\star$ as a function of the Kolmogorov exponent $x_K$ in both asymptotics $x_K \to 1^+$ and $x_K \gg 1$, namely :
\begin{equation}
 x_\star-1 \sim e^{\beta_0^\pm(n)} (x_K-1)^{\beta_1^\pm (n)} \text{   as  } \log (x_K-1) \to \pm \infty.
\label{eq:powerlaw}
\end{equation}
For the values of $n$ here considered, these power-law regimes seem to hold for  $x_K \ge x_K^+ = 1.4$ and $x_K \le x_K^- = 1.1$  (see Figure \ref{fig:engineeringfit}). Note that the behaviors of the $\beta^\pm_{0,1}(n)$'s (Figure \ref{fig:engineeringfit}) is compatible with the asymptotics described in the previous section. Indeed,  for large $n$ for example, we observe that $\beta_1^\pm \simeq 1$ and $\beta_0^\pm \simeq 0$, implying $x_\star \simeq x_K$ a feature indeed in agreement with the limit $(n\to\infty)$ previously described.

In a more practical perspective, we also wish to  get a ready-to-use proxy for the anomalous exponent $x_\star$ associated to the physical examples described in  Subsection \ref{ssec:leithexamples}. The behavior \ref{eq:powerlaw} suggests to model the \emph{logarithm} of $x_\star -1$ as a $n$-dependent piecewise linear function of $\log(x_K-1)$.
We therefore propose the following ``engineering fit'' :
\begin{equation}
\begin{split}
 &\log \left(x_\star-1 \right) =
\begin{cases}
&\beta_1^-(n) \log\left(x_K-1\right) + \beta^-_0(n)  \hspace{1cm} \text{if } x_K \le x_K^- = 1.1\\
&\beta_1^+(n) \log\left(x_K-1\right) + \beta^+_0(n)  \hspace{1cm} \text{if } x_K \ge x_K^+ = 1.4\\
&l_\star^- +  \dfrac{l_\star^+ - l_\star^-}{l_K^+ - l_K^-} \left(\log\left(x_K-1\right) - l_K^-\right)  \hspace{1cm} \text{otherwise,} \\
\end{cases}\\
& \text{with} \hspace{0.5cm} l_K^\pm = \log(x_K^\pm-1), \hspace{0.5cm} l_\star^\pm = \beta_1^\pm \log(x_K^\pm-1) + \beta_0^\pm,\\
&\hspace{1cm} \beta_0^\pm(n)= \exp \sum_{i=0}^4 \alpha_{0i}^\pm \log^i n, \hspace{0.5cm} \text{ and } \beta_1^\pm(n)= 1+ \exp \sum_{i=0}^4 \alpha_{1i}^\pm \log^i n.
\end{split}
\label{eq:engineeringfit}
\end{equation}
The $\alpha^\pm$'s are fitting coefficients whose values are given in Table \ref{table:engineeringfit}. These numbers reproduce the numerical data with less than $0.5$ percent of error on the value of $x_\star$ for the range of parameters here considered. However, the formula (\ref{eq:engineeringfit}) should not be extrapolated boldly outside its range of validity,  in particular in the regions of the parameter space where it is expected that the deviations from Kolmogorov regime are very large -- namely here $x_K>7$ or   $n <  1.05$.  The $4^{th}$ order polynomial fittings of the $\beta$'s indeed give a $n\to$ 0 behavior which disagrees with the asymptotics previously described.

\renewcommand*{\arraystretch}{1.5}
\begin{table}
\centering
\begin{minipage}{0.01\textwidth}
\centering
d=3
\end{minipage}
\begin{minipage}{0.98\textwidth}
\centering
\begin{tabular}{c|ccccc }
 & i=0 & i=1 & i=2 & i=3 & i=4   \tabularnewline
\hline
\hline
$\alpha_{0i}^-$ & $-1.14\cdot 10^{1}$ & $-1.35\cdot 10^{1}$  &  $-6.14\cdot 10^{0}$ & $-1.405\cdot 10^{0}$ & $-1.31\cdot 10^{-1}$\tabularnewline
$\alpha_{1i}^-$ & $-1.28\cdot 10^{1}$ & $-1.34\cdot 10^{1}$  &  $-6.05\cdot 10^{0}$ & $-1.36\cdot 10^{0}$ & $-1.27\cdot 10^{-1}$\tabularnewline
$\alpha_{0i}^+$ & $-2.79\cdot 10^{0}$ & $-2.92\cdot 10^{0}$  &  $-1.15\cdot 10^{0}$ & $-2.98\cdot 10^{-1}$ & $-3.48\cdot 10^{-2}$\tabularnewline
$\alpha_{1i}^+$ & $-2.83\cdot 10^{0}$ & $-1.78\cdot 10^{0}$  &  $-1.30\cdot 10^{0}$ & $-3.79\cdot 10^{-1}$ & $-4.93\cdot 10^{-2}$\tabularnewline
\end{tabular}
\end{minipage}\\
\begin{minipage}{0.01\textwidth}
\centering
d=2
\end{minipage}
\begin{minipage}{0.98\textwidth}
\centering
\begin{tabular}{c|ccccc }
\hline
\hline
$\alpha_{0i}^-$ & $-8.46\cdot 10^{0}$ & $-9.98\cdot 10^{0}$  &  $-4.26\cdot 10^{0}$ & $-9.02\cdot 10^{-1}$ & $-7.78\cdot 10^{-2}$\tabularnewline
$\alpha_{1i}^-$ & $-9.88\cdot 10^{0}$ & $-9.91\cdot 10^{0}$  &  $-4.20\cdot 10^{0}$ & $-8.61\cdot 10^{-1}$ & $-7.34\cdot 10^{-2}$\tabularnewline
$\alpha_{0i}^+$ & $-2.49\cdot 10^{0}$ & $-2.77\cdot 10^{0}$  &  $-1.21\cdot 10^{0}$ & $-3.46\cdot 10^{-1}$ & $-4.22\cdot 10^{-2}$\tabularnewline
$\alpha_{1i}^+$ & $-2.85\cdot 10^{0}$ & $-1.35\cdot 10^{0}$  &  $-1.16\cdot 10^{0}$ & $-3.59\cdot 10^{-1}$ & $-4.87\cdot 10^{-2}$\tabularnewline
\end{tabular}
\end{minipage}
\caption{Fitting constants used for the formula (\ref{eq:engineeringfit}) in the cases $d=3$ and $d=2$.}
\label{table:engineeringfit}
\end{table}

\section{The first-order model}
\label{sec:The first-order model}

One of the drawback of the generalized Leith models (\ref{eq:leithmodeldim}) is that we do not know whether the anomalous spectrum can be entirely and explicitly determined mathematically.  Let us now consider an even simpler model of turbulence based on a nonlinear first-order PDE:
\begin{equation}
\label{eq:kovazh}
\partial_t E = - C \partial_k (k^p E^q) - \nu k^2 E,
\end{equation}
where real (usually positive rational) constants $p$ and $q$ are chosen so that in the case of zero viscosity coefficient $\nu$ there is a power-law constant-flux solution of Kolmogorov type
$E \sim k^{-x_K}$, i.e.
$
p = q\, x_K.
$
Note that this model no longer has a thermodynamic equilibrium solution.
The second relation between $p$ and $q$ and the (usually dimensional) constant $C$ depends on the particular physical problem.

Let us mention again our main examples and specify the values $p$ and $q$ for the particular systems.

\subsection{Examples}

\subsubsection{3D hydrodynamic turbulence}

The  case with $ p=5/2$ and $q=3/2$ corresponds to the Kovasznay model  introduced in \cite{Kovasznay48}. In this case constant $C$ is dimensionless.

\subsubsection{Passive scalar turbulence}

In this case $E$ has the meaning of the passive scalar spectrum and, since the advection is passive, the resulting equation must be linear in $E$; therefore $q=1$.
The constant $p$ depends on the roughness of the advecting velocity field.
For Batchelor's smooth-velocity  regime  $p=1$ so that $x_K = 1$ (\cite{Batchelor59}).
For a rough velocity corresponding to the Kolmogorov spectrum (Obukhov-Corrsin regime) we have $p=5/3$ so that $x_K = 5/3$ (\cite{Obukhov49,Corrsin51}).

\subsubsection{Wave turbulence}

In this case  $q$ is the order of the resonant wave interaction less one, e.g. $q=2$ for three-wave processes, $q=3$ for four-wave processes, etc.
The constant $p$ depends on the particular type of the waves. Particular examples include:

\begin{enumerate}
\item
{\em Gravity water waves:} In this case  $q=3$ (a four-wave processes) and $p=15/2$ so that $x_K = 5/2$.
\item
{\em Capillary water waves:} In this case $q=2$ (a three-wave processes) and $p=7/2$
so that $x_K = 7/4$.
\item
{\em Sound  waves:} In this case $q=2$ (a three-wave processes) and $p=3$ so that $x_K = 3/2$.
\item
{\em  Alfven waves:} In this case $q=2$ (a three-wave processes) and $p=4$ so that $x_K = 2$.
\item
{\em Kelvin waves on vortex filaments:} In this case  $q=3$ (a four-wave processes) and $p=5$ so that $x_K = 5/3$.

\end{enumerate}

\subsection{Analysis of the first-order model}

Let us make in equation (\ref{eq:kovazh}) the  change of variables
$E = \varepsilon^{1/q}  k^{-p/q} $ ($\varepsilon$ is the energy flux). We have:
\begin{equation}
\label{eq:kovazh0}
\frac 1 q    \partial_t \varepsilon = - C k^{p/q}  \varepsilon^{1-1/q} \partial_k \varepsilon - \nu  k^{2} \varepsilon.
\end{equation}

\subsubsection{{Hydrodynamic and wave turbulence (Case $q>1$)}.}

 First, let us consider the case $q > 1$.
Making a further change of variables
\begin{equation}
\label{eq:kovazh2}
u =  \varepsilon^{1-1/q} \quad \hbox{and}
\quad
%\ell = k^{1-p/q},
{\ell = k^{1-p/q}},
\end{equation}
we have
\begin{equation}
\label{eq:kovazh1}
   \partial_t u =  \tilde C  \,  \partial_\ell \left( u^2 \right) - \nu   (q-1)  \ell^{2/(1-p/q)} u,
\end{equation}
where
$
   \tilde C =  {C (p-q)}/ 2.
$
%For $x_K > 1$ (which is the case in all our examples with $q \ne 1$) we have
%$\tilde C >0.$
{In all our examples with $q \ne 1$, we have $1<q <p < 3q$. We later assume that this relation holds. This implies $\tilde C >0$, and $x_K > 1$ so that $p/q-1 >0$ in the aforementioned change of variables (\ref{eq:kovazh2}) (recall that $x_K =p/q$).} The upper bound on $p$ will be useful to describe the dynamics after the shock.

It is easy to see that in the case $\nu =0$,  we recover an inviscid 1D Burgers equation.
Thus, at the initial evolution stage when the viscosity effect is negligible we should
expect features of the Burgers behaviour, in particular  wave breaking leading to a pre-shock singularity. Thereafter, we can also anticipate a shock formation. However, the viscous term in equation  (\ref{eq:kovazh1}) is obviously different from the one of the Burgers equation and, therefore, the shock structure should be carefully re-examined.

Starting with the inviscid stage, we first note that (unlike the case of the second-order models) the initial conditions  for the spectrum $E$ with finite  support in $k$ are not
appropriate: if the velocity at the boundaries of the support is zero then the support will not grow in time and there will be no cascade to larger $k$'s. Thus we will consider the initial spectrum which extends to infinite $k$ (although decreasing there as will be discussed later). For simplicity, let there be a minimal wave number $k_{min}$ below which the initial spectrum is zero.
  General smooth profiles like this have both positive and negative derivative parts, and their evolution will lead to wave breaking  with an infinite derivative, \emph{ie} $\partial_\ell u = \infty$, at a single point $\ell = \ell_*$ at a time $t=t_\star$. Since $ \tilde C >0$, the breaking will happen somewhere where $\partial_\ell u>0$. We will suppose that the initial profile $u_0(\ell)$ is such that in the range of positive  $\partial_\ell u_0(\ell)$ the second derivative is negative,
$\partial_{\ell \ell} u_0(\ell) <0$. In this case the breaking occurs at $l_*=0$. Provided that $\partial_{\ell} u_0(0) \ne0$ (this corresponds to $E(k) \sim k^{-(p-1)/(q-1)}$ at large $k$ initially), we have $u(\ell) \sim \ell^{1/3}$ at time $t=t_\star$ in the vicinity of the breaking point for the profile. In terms of the spectrum, such a behaviour implies
\begin{equation}
\label{eq:Eburg}
E(k) = u^{\frac{1}{q-1}}  k^{-p/q} \sim
\ell^{\frac{1}{3(q-1)}-\frac{p}{ q-p}} = k^{-x_\star}.
\end{equation}
where
\begin{equation}
\label{eq:EburgX*}
x_\star = {\frac{p-q}{3q(q-1)}+\frac{p}{ q}}.
\end{equation}
Note that this spectrum is steeper than the Kolmogorov-type spectrum,
$x_\star  > x_K =p/q$. Again, we see that the limits $p \to q$ and $q \to 1$ do not commute. Recall that the limiting limit,  $p = q$ corresponds to the passive scalar turbulence in smooth velocity field---a (marginally) infinite capacity system which is considered later in this section.
%
%For example, for the Leith model $x_\star =2.1111$ and for the MHD waves $x_\star =7/3$.  The latter value coincides with the value obtained by numerical simulations of the kinetic equation in  \cite{galtier2000weak}.
%
To obtain, the large $k$ behavior of the spectrum, we take into account $
u_0(\ell)  = (k^{p/q} E)^{q-1}
$ and
$
%\partial \ell = (1-p/q) k^{-p/q} \partial k, \quad
\partial_\ell = \frac {k^{p/q}} {(1-p/q)}  \partial_k
$.
 The condition  $\partial_{\ell \ell} u_0(\ell) <0$ then reads:
$$
%\frac 1{(1-p/q)^2} k^{p/q}
\partial_k \left( k^{p/q} \partial_k (k^{p/q} E)^{q-1} \right) <0.
$$
Condition $\partial_{\ell} u_0(0) \ne 0$ and the condition that the flux is zero at $k =\infty$
leads to the asymptotic behaviour  $u_0(\ell) \sim \ell$ at $\ell \to 0$.
This implies a power-law spectrum $E \sim k^{-x} $ at  $k \to \infty$
with $
x=\frac{p-1}{q-1}.
$

\begin{comment}
 this means
$$
%\partial_k \left( {(p/q -x)(q-1)}   k^{(p/q -x)(q-1) -1+p/q} \right) =
 {(p/q -x)(q-1)}   [{(p/q -x)(q-1) -1+p/q} ] <0.
$$
Condition $\partial_{\ell} u_0(\ell) > 0$ means the flux decaying with $k$.
For a power-law spectrum $E \sim k^{-x} $ this means
$$x> p/q  .$$
Taking this condition and the condition $q>1$ into account, for the condition
 $\partial_{\ell \ell} u_0(\ell) <0$ we have:
$$
   {(p/q -x)(q-1) -1+p/q} =
% {p -xq -p/q +x -1+p/q}  =
 p -xq  +x -1
>0,
$$
or
$$
x< \frac{p-1}{q-1}.
$$

$
E = u_0(\ell)^{1/(q-1)}  k^{-p/q} .
$
Let $u_0(\ell) \sim \ell = k^{1-p/q}$:
$$
E = k^{(1-p/q)/(q-1) -p/q}  =
k^{(1-p)/(q-1) }
.
$$
\end{comment}
%

For example, the anomalous transient exponent is $x_\star =2.1111$ for 3D hydrodynamics and $x_\star =7/3$ for the MHD values.  Note that the latter value coincides with the value obtained by numerical simulations of the kinetic equation in  \cite{galtier2000weak}.
In both cases,  we  obtain
$ x= 3$.

%Actually, the same behaviour will for initial spectra with $x<\frac{p-1}{q-1},$ provided the $x>x^*$.

Now let us consider dynamics after the spectral front reaches the dissipative scale, $t>t_\star$.
This will be characterised by a shock in the profile $u(\ell)$ near $\ell=0$  the structure of which is determined by the dissipation term in equation (\ref{eq:kovazh1}). Within the shock one can neglect the time derivative term :
\begin{equation*}
\label{eq:kovazh1nu}
 2  \tilde C  \,  \partial_\ell  u  - \nu   (q-1)  \ell^{2/(1-p/q)} =0.
\end{equation*}
Solving this equation and taking into account the condition $u(0) =0$ yields :
\begin{equation*}
\label{eq:kovazh2nu}
    u(\ell) =  C_1 - C_2 \ell^{\frac{3q-p}{q-p} }, ~\text{ with }~
%\end{equation}
%where
%\begin{equation}
%\label{eq:kovazh2nu1}
    C_2 =% - \frac{ \nu   (q-1)(q-p)}{  2 \tilde C (3q-p)} =
\frac{  \nu   (q-1)}{    C (3q-p)}  >0
\end{equation*}
%(we will assume $p<3q$) and
{(recall that $p<3q$ is here assumed)}.
$C_1 >1$ is independent of $\ell$ but may be dependent on $t$.
The first term on the right-hand side here is negative and singular at $\ell =0$.
Therefore,   the dissipation makes the spectrum  have a cut-off wave number $k_{\nu}$ corresponding to some
$\ell = \ell_\nu $ such that $ u(\ell_\nu)=0$, i.e.
\begin{equation}
\label{eq:kovazh3nu}
   \ell_\nu=  \left[\frac{   C_1  C (3q-p)} {  \nu   (q-1)}
\right]^{\frac{q-p}{3q-p} } \quad \hbox{and} \quad
%k_\nu=  \left[\frac{   C_1  C (3q-p)} {  \nu   (q-1)}
%\right]^{\frac{(q-p)^2}{q(3q-p)}}.
{k_\nu=  \left[\frac{   C_1  C (3q-p)} {  \nu   (q-1)}
\right]^{\frac{q}{3q-p}}}.
\end{equation}
The value of $C_1$ is to be determined from the matching to the jump in $u(\ell)$ arising from the inviscid Burgers solution. In the vicinity of the wave breaking point, the velocity profile behaves as
$\ell + u (t-t^*) \sim u^3  $, so that
 $C_1 \sim \sqrt {t-t^*}$  for the jump at $\ell =0$.
For the spectrum, this implies :
\begin{equation}
\label{eq:kovazh2nu2}
    E(k) = \left( C_1 -C_2 k^{{3-p/q}} \right)^{\frac 1 {q-1}} k^{-\frac pq}.
\end{equation}
For $k<k_\nu$ the spectrum is of Kolmogorov type and satisfies :
$  E(k) =  C_1^{\frac 1 {q-1}} k^{-\frac pq}
\sim (t-t^*)^{\frac 1 {2(q-1)}} k^{-\frac pq},
$. We  therefore see that the Kolmogorov-type spectrum invades the $k$-space propagating from large to low $k$'s. It  gradually replaces the anomalous spectrum $E \sim k^{-x_\star}$ (whose amplitude is almost time-independent).

\subsubsection{Passive scalar turbulence (Case $q=1$)}
Let us now consider the case $q=1$, $p > 1$, which corresponds to a passive scalar in a rough velocity field. The case can be treated by the same method and similar results are found. As  before, there exist
no power-law asymptotics is formed before the dissipative scale is reached. The equation for the energy flux now reads :
\begin{equation}
%\label{eq:kovazh0sm1}
   \partial_t \varepsilon = - Ck^p  \partial_k \varepsilon - \nu  k^{2} \varepsilon.
\end{equation}
Using the further change of variables $\ell = k^{1-p}$,
%\begin{equation}
%\label{eq:kovazh2smm}
%\ell = k^{1-p},
%\end{equation}
it becomes
\begin{equation*}
\label{eq:kovazh0sm111}
   \partial_t \varepsilon = (p-1) C \partial_k \varepsilon - \nu  \ell^\frac{2}{1-p} \varepsilon.
\end{equation*}

Writing $\varepsilon (k,t) = F(\kappa, t) $ with $\kappa = \ell +(p-1) Ct$,  we obtain
\begin{equation*}
%\label{eq:kovazh1sm}
   \partial_t F = - \nu  (\kappa -(p-1)  Ct)^\frac{2}{1-p} F,
%\end{equation*}
%from which it follows that :
%\begin{equation*}
%\label{eq:kovazh2sm}
~ \text{hence } ~
   \ln F = \frac{ \nu}{ (p-3)  C}  (\kappa -(p-1)  Ct)^\frac{3-p}{1-p} + G(\kappa),
\end{equation*}
with $G$ being an arbitrary function.
Plugging the original variables back in yields :
\begin{equation}
\label{eq:kovazh3sm}
%  E(k,t) = k^{-p} e^{\frac{ \nu}{ (p-3)  C}  k^{3-p}  } \tilde G(k^{1-p}+(p-1)  Ct),
  E(k,t) = k^{-p} e^{-\frac{ \nu}{ (3-p)  C}  k^{3-p}  } \tilde G(k^{1-p}+(p-1)  Ct),
\end{equation}
with  $\tilde G(k) = e^{G(k)} $ -- a function defined by the initial condition.
For the finite capacity case ($p>1$), the initial spectrum propagates toward low values of $\ell = k^{1-p}$, therefore towards high $k$'s. The propagation  speed is constant in the space of variable $\ell$. This  means that the front reaches $\ell=0$ (or alternatively $k=\infty$) in a finite time. Therefore, for any viscosity $\nu$, no matter how low, the dissipative scale
\begin{equation}
{k_\nu = \left( \frac{ 3-p  }{ \nu} C \right)^{\frac{1}{3-p}}}
\end{equation}
is reached in a finite time $t_\star \approx k_0^{1-p} /(p-1)  C$.

For $t < t_\star$, the evolution is inviscid and non-universal: the spectral  slope has Kolmogorov value $-p$ near the maximum of function
$\tilde G(k)$, but it gradually varies as one moves away from this maximum.
At $t \sim t_\star$, the maximum of
$\tilde G(k)$ reaches $k_\nu$ and stays there thereafter, gradually becoming flatter on its left side, $k <k_\nu$.
This corresponds to spreading of the Kolmogorov scaling $E \sim k^{-p}$ from the dissipative scale $k_\nu$ into the inviscid range $k <k_\nu$ ---a picture we have anticipated.

 Let us now consider the case  $q=p=1$, which corresponds to the example of the passive scalar turbulence in a smooth velocity field.
The energy flux satisfies
\begin{equation}
\label{eq:kovazh0sm}
   \partial_t \varepsilon = - C k \partial_k \varepsilon - \nu  k^{2} \varepsilon.
\end{equation}
Let $\varepsilon (k,t) = F(\sigma, t) $ with
 $\sigma(k,t)= Ck \exp(-Ct)$. We obtain :
\begin{equation}
\partial_t \log F = - \nu \dfrac{\sigma^2}{C^2}\exp(2Ct),
%\end{equation}
~\text{so that}~
%\begin{equation}
F(\sigma,t) = G(\sigma)\exp^{-\dfrac{\nu}{2C^3} \sigma^2e ^{2Ct}}.
\end{equation}
The function $G(\sigma)$ has to be fixed by the initial condition
$E(k,0) = E_0(k)$. This gives:
\begin{equation}
E(k,t) = E_0(ke^{-Ct}) \, \exp \left[ {-Ct-\dfrac{\nu k^2}{2C}(1- e^{-Ct} )}\right].
\end{equation}

From this solution we see that the inviscid stage lasts until the time
$\sim 1/C$ when the front reaches the dissipative wave number
$k_\nu = \sqrt{2D/\nu}$. At this stage the spectrum is simply being stretched
exponentially in time to larger $k$'s in a self-similar way conserving the
total energy.  For $t > 1/C$, this stretching continues in the range
$k < k_\nu $ getting more and more flat, whereas at $k \sim k_\nu $ the spectrum experiences
an effective cut-off.
Note that at no point one observes the Kolmogorov-type scaling, which is a natural behaviour  for the free decay in infinite capacity systems.

\section{Numerics}
\changed{In order to illustrate the previous analysis, we now take the example of Alfven wave turbulence, and show the outcomes of numerical simulations of both modelings: first and second-first-orderorder. In principle,  the first-order model is fully integrable by the method of characteristics. However, the resulting solutions are typically implicit and numerical solutions remain helpful to visualize the evolution.}

\begin{figure}
\centering
\begin{minipage}{0.64\textwidth}
\includegraphics[width=\textwidth]{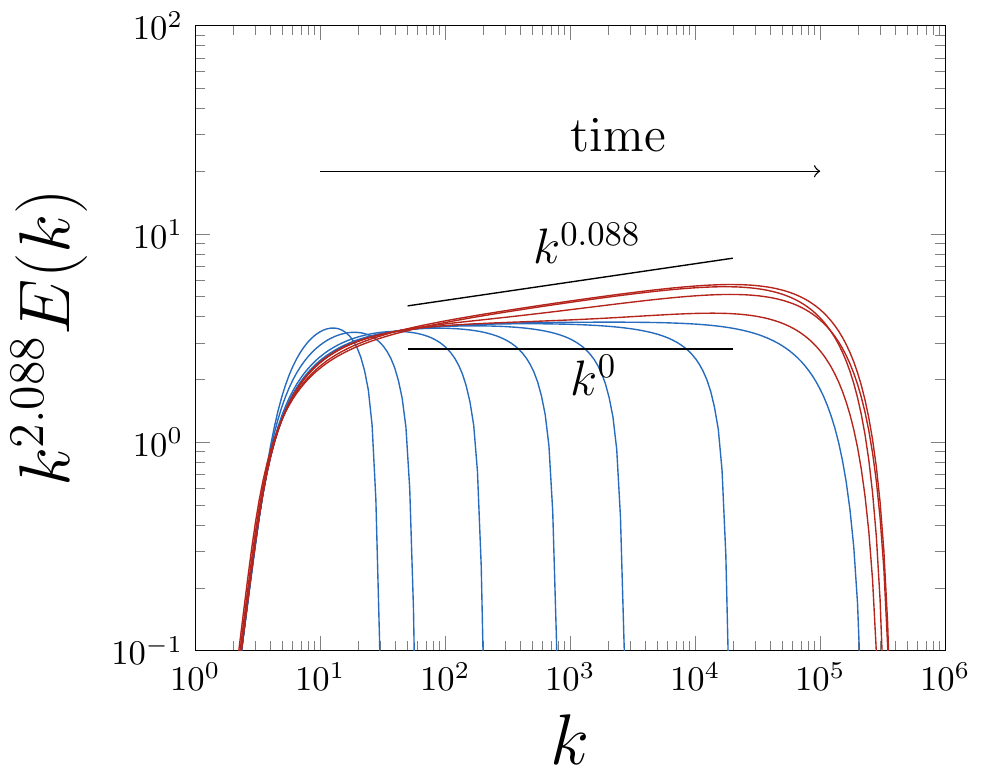}
\end{minipage}
\begin{minipage}{0.35\textwidth}
\includegraphics[width=\textwidth]{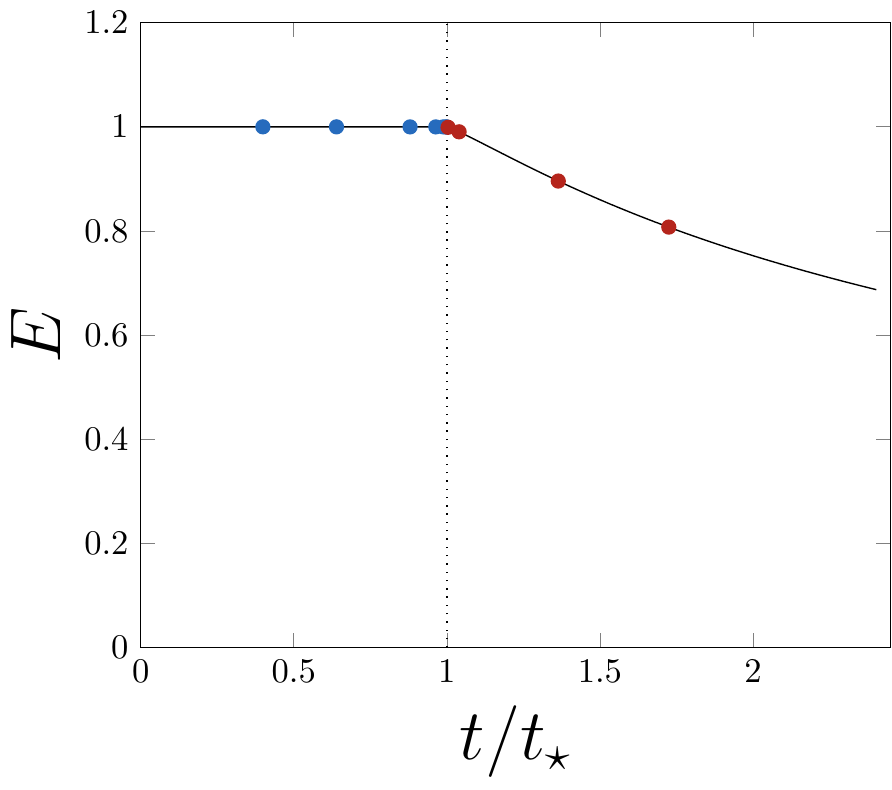}
\end{minipage}
 \caption{Left: Compensated MHD spectra  obtained numerically from the second-order differential model at different times: earlier than $t_\star$ (in blue) and later than $t_\star$ (in red). The spectra are normalized by the total energy. The times are  reported by the dots on the right inset. The blue dots correspond to $t/t_\star-1\simeq -6 \cdot 10^{-1}$, $-4 \cdot 10^{-1}$, $-10 ^{-1}$, $-4 \cdot 10 ^{-2}$, $-10 ^{-2}$, $-2\cdot 10 ^{-3}$, $0$; the red dots to $t/t_\star-1\simeq 4 \cdot 10^{-3}$, $4 \cdot 10^{-2}$, $-10 ^{-1}$, $4 \cdot 10 ^{-1}$, $7 \cdot10 ^{-2}$.  The right inset shows the time evolution of the energy. $t_\star \simeq 8.3 \cdot 10^{-3}$ marks the onset of energy dissipation.  The figure is made using the data obtained by E.Buchlin and previously reported in \cite{galtier2010nonlinear} (in a different form).}
 \label{fig:LeithMHD}
\end{figure}

\changed{The details of the numerical simulations are the following.
For the second-order model, we solve  Equation \ref{eq:leithmodeldim} with $d=2$, $n=1$ and $m=6$ and a small viscosity $\nu=5 \cdot 10^{-5}$ as performed in \cite{galtier2010nonlinear}. The initial spectrum is $E(k,0) \propto k^3 e^{-k^2/25}$, such that $E(t=0)=1$. The dynamics is then evolved until  time $t_{f}=0.02$ using a Crank-Nicolson scheme with adaptative time-stepping. $200$ values of '$k$'s are chosen, logarithmically spaced from $k=1$ to $ k=2^{20}$.
For the first-order model,  we use Equation (\ref{eq:kovazh1}) with $q=2, p=4$ and add a very small viscosity, namely
$\nu = 5 \cdot 10^{-4}$ for a resolution of $4,000$ points.
As an initial condition, we chose $u_0(\ell) = \sin(\ell/4)$ for $0 \le \ell \le 4 \pi$ and $u_0(\ell) =0$ for $\ell > 4 \pi$. This initial profile satisfies  $\partial_{\ell \ell} u_0(\ell) <0$ in the range of positive  $\partial_\ell u_0(\ell)$ together with  $\partial_{\ell} u_0(0) \ne0$. It therefore matches the conditions of the previous analysis. Crank-Nicolson and Adam-Bashforth  numerical schemes are used to reach the final time $t_f=2000$, choosing $dt=10^{-2}$ for the time stepping. Note that the initial spectrum at large $k$  behaves as $E(k) \sim k^{-(p-1)/(q-1)} = k^{-3}$ which is steeper than both the Kolmogorov-type spectrum and the transient anomalous spectrum (see Figure  \ref{fig:sp1st}).}
\changed{The results for the second-order model are shown on Figure  \ref{fig:LeithMHD}.
For this example, recall that the anomalous exponent is $x_\star \simeq 2.088$ while the Kolmogorov exponent is $x_K =2$. The difference between the anomalous transient exponent and the Kolmogorov exponent is therefore very small, less than 5 percent. However, this difference is clearly apparent in our numerics. Following the evolution of the total energy, one can identify $t_\star \simeq 8.3 \cdot 10^{-3}$ as the onset of energy dissipation (see the right panel of Figure  \ref{fig:LeithMHD}).
 For times earlier than $t_\star$,  the figure clearly displays an  anomalous growth of the energy spectrum as $k^{-x_\star}$
with $x_\star \simeq 2.088$ as predicted by our analysis.
After the spectrum reaches the dissipative scale, it is gradually replaced by the Kolmogorov-type spectrum $E(k) \sim k^{-2}$ (so that the compensated spectrum behaves as : $k^{x_\star} E(k) \sim k^{0.088}$). This numerical experiment provides a clear illustration of the calculations presented in section \ref{sec:Self-similar solutions of the nonlinear diffusion models}.}

\begin{figure}
 \centering
\begin{minipage}{0.49\textwidth}
\includegraphics[width=\columnwidth]{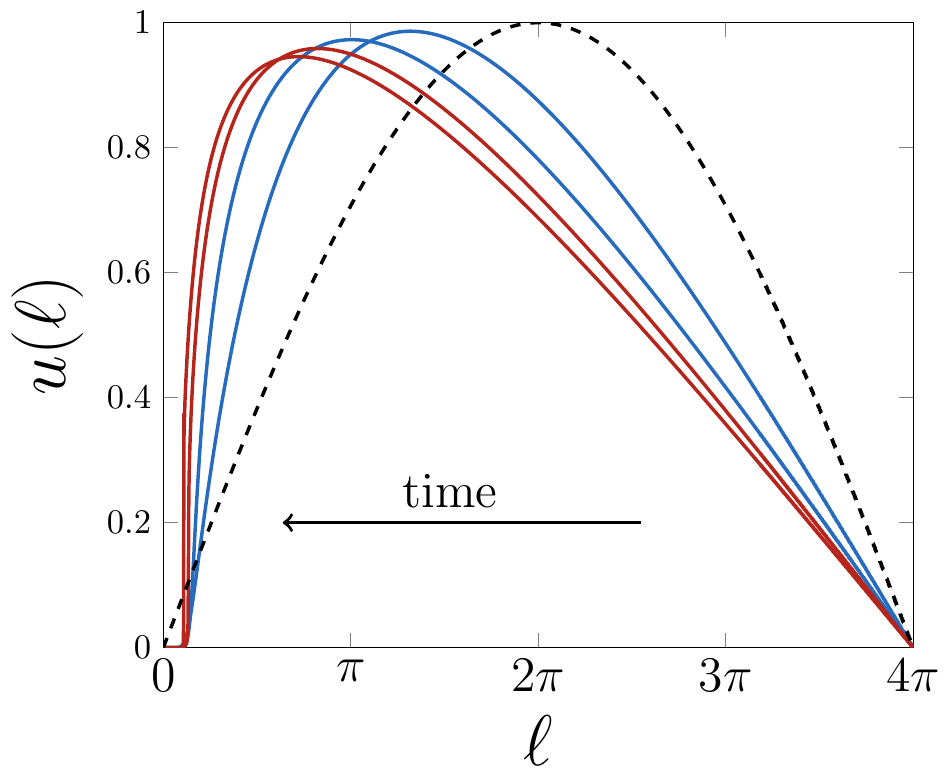}
\end{minipage}
\begin{minipage}{0.49\textwidth}
\includegraphics[width=\columnwidth]{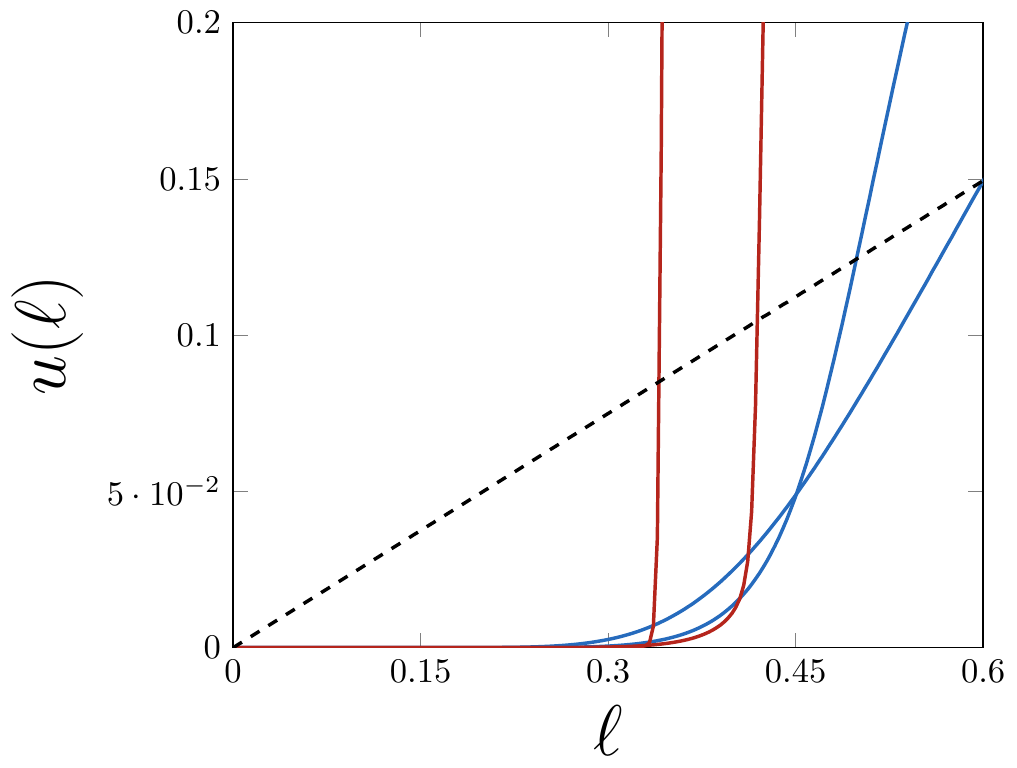}
\end{minipage}
 \caption{Left panel : function $u(\ell, t)$ for the first-order differential model of MHD turbulence  at different moments of time. Right panel: zoom at the small-$\ell$ region.
 The initial profile is in dashed black.  The times before the shock are  $t=80$, $400$ (in blue). The times after the shock are  $t=640$, $800$ (in red). Please see also the inset on the right of Figure  \ref{fig:sp1st}. }
 \label{fig:u1st}
\end{figure}

\begin{figure}
\centering
\includegraphics[width=0.49\columnwidth]{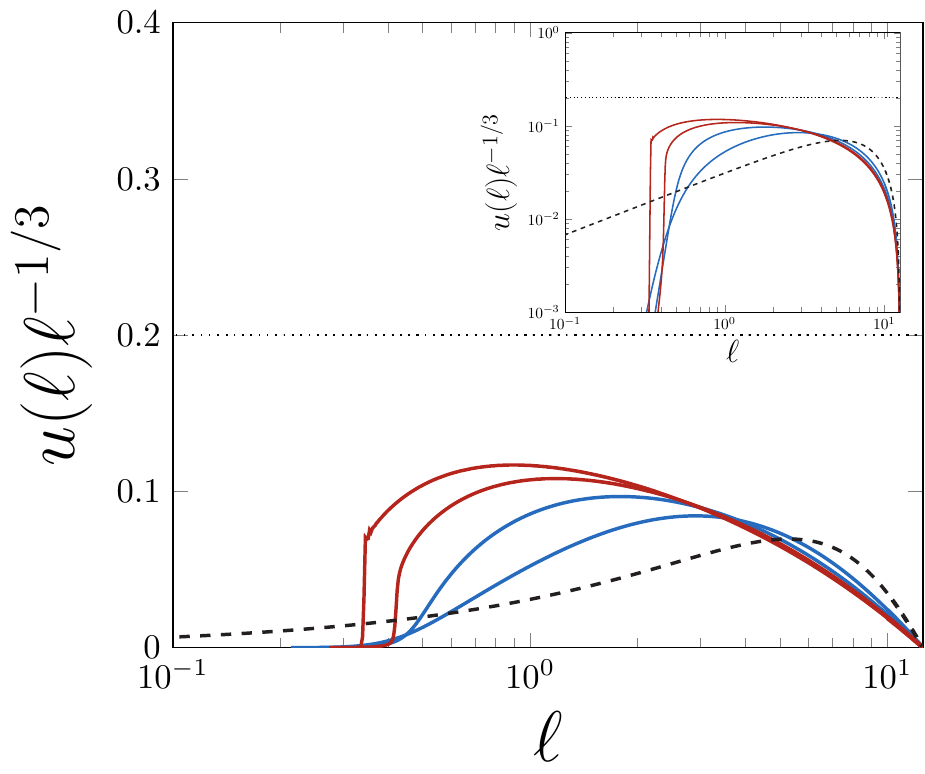}
\includegraphics[width=0.49\columnwidth]{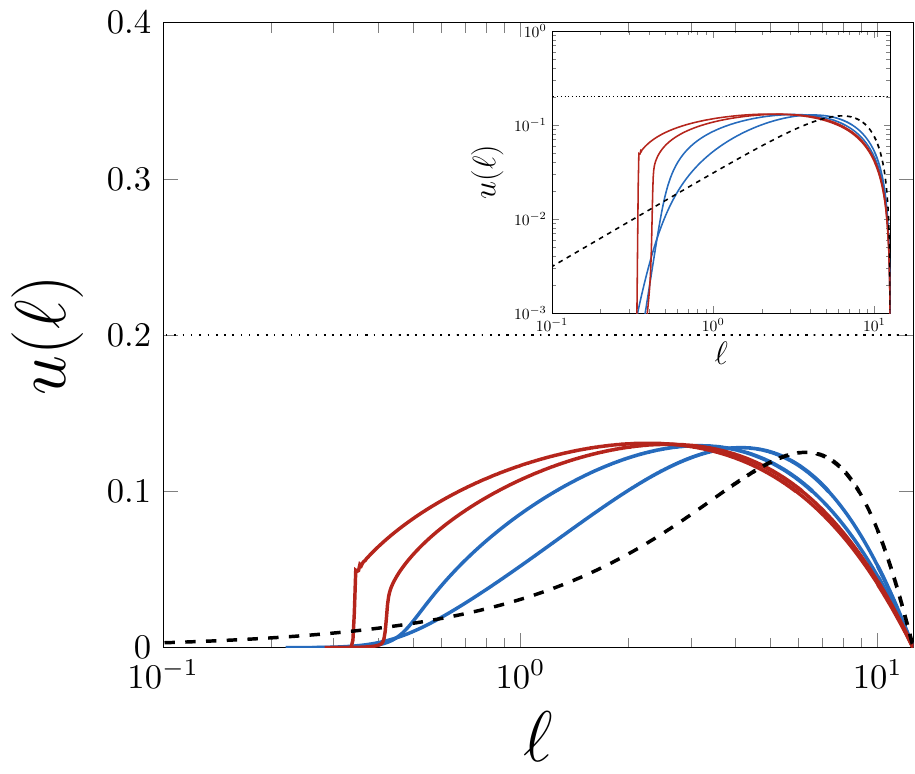}
 \caption{Compensated velocity profiles  for  the first-order differential model at different times.  The spectra are  compensated by $l^{1/3}$ (left panel) and $l^{0}$ (right panel). The times  displayed are those of Figures \ref{fig:u1st}. The insets show the same data using log-log coordinates.}
 \label{fig:u2st}
\end{figure}

\begin{figure}
\centering
\begin{minipage}{0.59\textwidth}
\includegraphics[width=\textwidth]{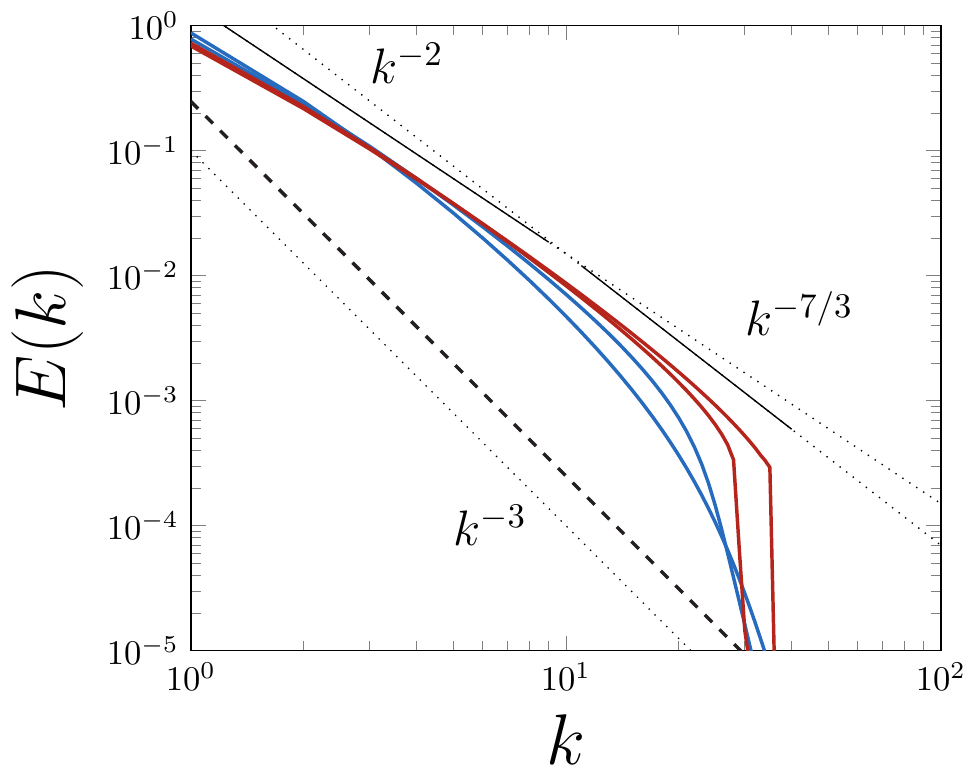}
\end{minipage}
\begin{minipage}{0.35\textwidth}
\includegraphics[width=\textwidth]{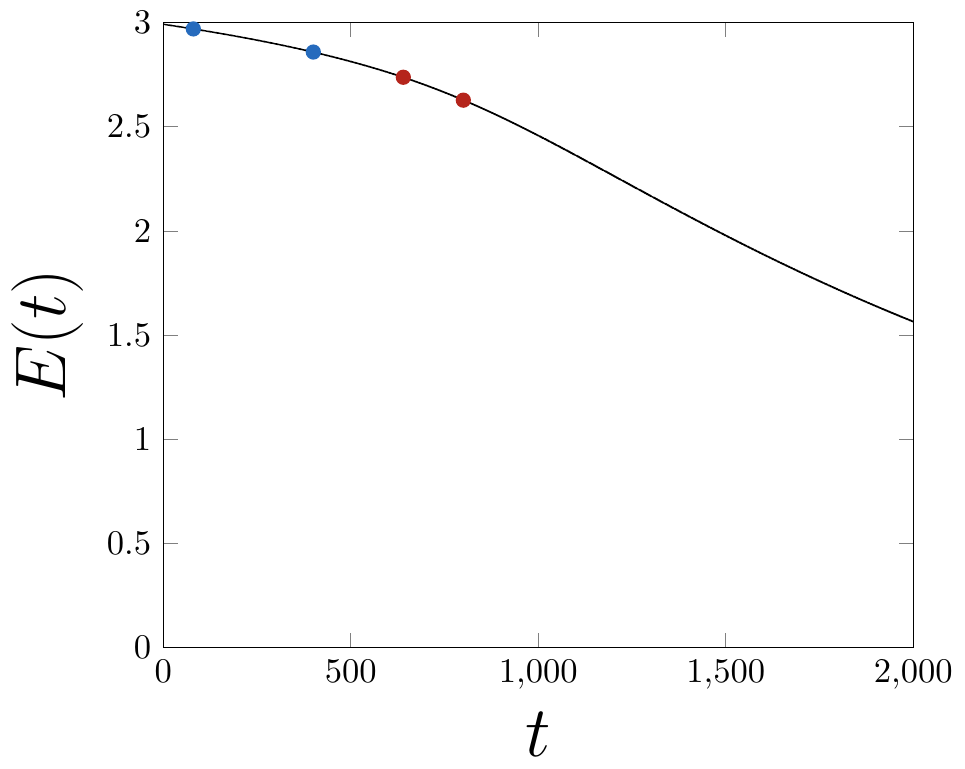}
\end{minipage}
 \caption{Spectrum of MHD turbulence obtained numerically from  the first-order differential model at different moments of times. The times and the color legend are the same as in Figure  \ref{fig:u1st} . The right inset shows the time evolution of the total energy, with the blue and red dots indicating the times picked  in Figures \ref{fig:u1st} to  \ref{fig:sp1st}.}
 \label{fig:sp1st}
\end{figure}

\changed{The results for the first-order model are shown on Figures \ref{fig:u1st},  \ref{fig:u2st} and \ref{fig:sp1st}.
The time evolution of the  profile $u(\ell, t)$ is shown on
Figure \ref{fig:u1st}. The figure displays the salient feature of the first-order model, namely  the formation of a sharp cut-off at small
$\ell$ through a wave breaking process. The compensated plots of In figures \ref{fig:u2st}  hint at the dominance of the $\ell^{1/3}$ inertial range scaling at the times close to $t^\star$, and  $\ell^{0}$ inertial range scaling at later times.
In terms of the spectrum, this should translate into the development of a transient spectrum
with index $x_\star =7/3$,  gradually replaced with a Kolmogorov-type spectrum  with index $x =2$ at later times.  Such a trend is detectable on  the spectra shown on Figure \ref{fig:sp1st}, corresponding to different times chosen both  before and after the front has developed.}
%\begin{figure}
% \centering%
%\includegraphics[width=0.49\columnwidth]{gu4-corr}
%\includegraphics[width=0.49\columnwidth]{gu3-corr}
% \caption{Log-log plots of function $u(\ell, t)$ for the first-order differential model of MHD turbulence  at different moments of time.  Left panel:  $u(\ell, t)$ compensated by $\ell^{1/3}$. Right panel : non-compensated plots. }
% \label{fig:u2st}
%\end{figure}
%

\section{Conclusions}
\label{sec:summary}

In the present paper, we have analysed  differential models for various examples of hydrodynamic, passive-scalar and wave turbulence given by  the second-order Leith-type nonlinear diffusion  and the first-order Kovasznay-type nonlinear transport equations.
Both types of models predict formation an anomalous (steeper than Kolmogorov) transient power-law spectra.

The second-order models were analysed in terms of self-similar solutions of the second kind, and the anomalous exponent was found numerically and presented via a phenomenological ``engineering fit" formula for a broad range of parameters relevant for the physical examples, both the known ones and potentially the new ones.
A peculiar non-commuting limit was noted for the systems close to the example of a passive scalar in a smooth velocity field.

The first-order models were examined analytically, including finding an analytical prediction for the anomalous exponent of the transient spectrum and description of formation of the Kolmogorov-type spectrum as a reflection wave from the dissipative scale back into the inertial range. Even though the first-order models are less realistic than the second-order models, their advantage is that they admit a full analytical treatment, which allows to relate the anomalous transient scaling and the subsequent
Kolmogorov-type spectrum to the pre-shock/shock singularities similar to the ones arising in the Burgers equation. \changed{On the other hand, somewhat counter-intuitively, the first-order model is much harder than the second-order model to deal with numerically. This is due to the  appearance of shocks, as typical for hyperbolic systems.}

It is tempting to think that in more realistic settings, including the second-order and the integro-differential models (e.g. EDQNM, kinetic equations), a similar link between the anomalous transient scaling and a singularity could exist. Moreover, such a link may  exist also on the dynamical level beyond the closures. With a bit of imagination one could
suggest that the long-hunted for Euler singularity, if exists, may be responsible for the anomalous scaling which is a precursor to the Kolmogorov spectrum. Clearly, at this moment in time such suggestion is a pure speculation. However, we would like to recall numerical simulations of
\cite{Brachet-pancakes}
(and more recently
\cite{Cichowlas2005239}) where a precursor power-law with a steep exponent $\sim 4$ was found for the hydrodynamic turbulence, and it was linked to transient pancake-like quasi-singular structures.
It would be interesting to study this effect numerically further using modern high-resolution codes, and also check the scenario of formation of the Kolmogorov spectrum as a backscatter wave in the Navier-Stokes turbulence.

Our final remarks are about the degree of universality of the transient anomalous exponents. On one hand, these exponents are universal in the sense that they are insensitive to the initial spectrum. However, clearly the exponents are model-dependent, e.g. they are different for the first-order and the second-order models of the same physical systems.
For example, for the Navier-Stokes turbulence $x_\star \approx 2.111$ for the first-order model, $x_\star \approx 1.851$ for the second-order model,
$x_\star \approx 2$ for the EDQNM closure, and $x_\star \sim 4$ in DNS.
Thus, the robust feature common to the different models is the fact that the anomalous scaling exist, but not the predicted value of its exponent.

\renewcommand*{\arraystretch}{1.5}
\begin{table}
\centering
\begin{tabular}{c|ccc }
 Model & Kolmogorov & Leith & Kovasznay    \tabularnewline
\hline
\hline
 3D hydrodynamics & $5/3$ & $\simeq 1.851$ & $19/9 \simeq 2.111$  \tabularnewline
 Passive Scalar (smooth) & 1 & undefined & undefined  \tabularnewline
 Passive Scalar (rough) & $5/3$ & $\infty$ & $\infty$  \tabularnewline
 Gravity Waves & $5/2$ & $\simeq 2.514$ & $11/4 = 2.75 $  \tabularnewline
 Capillary Waves & $7/4$ & $\simeq 1.799$ & $2$  \tabularnewline
 Sound Waves & $3/2$ & $\simeq 1.509$ & $5/3 \simeq 1.667 $  \tabularnewline
 Alfven Waves & $2$ & $\simeq 2.088$ & $7/3  \simeq 2.333 $   \tabularnewline
 Kelvin Waves & $5/3$ & $\simeq 1.672$ & $16/9 \simeq 1.778$  \tabularnewline
\end{tabular}
\caption{Anomalous exponents obtained for various turbulent systems described in terms of the (generalized) Leith and Kovasznay described in the paper.  }
\label{table:summary}
\end{table}

\section{Acknowledgements}

We acknowledge Eric Buchlin for helpful discussion.
Sergey Nazarenko gratefully acknowledges support of grant ``Chaire Senior PALM TurbOndes'' and hospitality of the SPEC lab, CEA, Saclay.

\appendix

\section{ Stability analysis of the fixed points}
\label{ap:stability}
The stability of the fixed points $P_1$, $P_2$ and $P_3$ defined by Equation (\ref{eq:steadydim})  can be determined by computing the Jacobian matrix $\Delta(\rho,\sigma)$ associated to  the system (\ref{eq:evodim1}) --  (\ref{eq:evodim2}). The latter reads
\begin{equation}
%\Delta(\rho,\sigma)=
\begin{pmatrix}
 \; \;  \; \;  \dfrac{\lambda}{\mu}(2\rho+\sigma) & \dfrac{\lambda}{\mu}\rho \\
 \dfrac{\mu x}{ \lambda(x-\lambda)}+2(d-1) \! \left[\dfrac{1}{\lambda}+ \dfrac{1}{\mu} \right] \rho + \! \left[\dfrac{d-1}{\mu}+ d-2 \right] \! \sigma & \dfrac{\mu}{x-\lambda} - 2\lambda\sigma {+} \! \left[ \dfrac{d-1}{\mu}+d-2\right] \! \rho
\end{pmatrix}
\nonumber
\end{equation}
In particular, at the fixed points $P_1, P_2$ and $P_3$, the Jacobian matrix respectively takes the values
\begin{equation}
\begin{split}
&\Delta_1=\dfrac{\mu}{ x-\lambda} \begin{pmatrix}
 0 & 0 \\
 x/\lambda & 1
\end{pmatrix}, \sieq
 \Delta_2 = \dfrac{\mu}{x-\lambda}
\begin{pmatrix}
 1 & 0 \\
 \dfrac{ x+ d-2 }{\lambda}  + \dfrac{ d-1}{\lambda\mu}& -1
\end{pmatrix} \sieq \text{and} \\
& \Delta_3 =
\begin{pmatrix}
 \dfrac{\lambda}{\mu}\rho_3 & \dfrac{\lambda}{\mu}\rho_3 \\
 \dfrac{\mu x}{ \lambda(x-\lambda)}+ \left[\left(d-1\right) \left(\dfrac{2}{\lambda}+\dfrac{1}{\mu}-1\right)+ 1 \right] \rho_3  & \dfrac{\mu}{x-\lambda} + \left[ 2\lambda -1+  (d-1)\left(1+\dfrac{1}{\mu}\right) \right]\rho_3
\end{pmatrix} \\
& \sieq \text{with $\rho_3$ defined as } \sieq  \rho_3 = \dfrac{\mu}{(\lambda-1)(\lambda+d-1)}.\end{split}
\end{equation}
For all known examples $\mu =1/n >0$. Then $P_1$ is a saddle-node and $P_2$ is a saddle.  To determine the nature of $P_3$, one needs to determine the eigenvalues of $\Delta_3$. We compute the trace $T_3$ and the determinant $D_3$ of $\Delta_3$ as:
\begin{equation}
 T_3=  \dfrac{\mu}{x-\lambda} + \left(\lambda\left(2+\dfrac{1}{\mu} \right) + (d-1)\left(1+ \dfrac{1}{\mu}\right) -1 \right) \rho_3 \sieq \text{and} \sieq D_3 = \rho_3.
\end{equation}
The determinant  $D_3$ is positive. Indeed, from the definitions of $x_K$ on one hand and of $(\lambda,\mu)$ on the other hand (Equation  (\ref{eq:lambdamu})), we obtain that  $\lambda = (\mu+1)x_K -\mu$, with $\mu =1/n >0$ and $x_K >1$ for the finite capacity scenarios that we here  consider. As the dimension $d$ is greater than $1$, we conclude that both $\lambda$, $\lambda +d-1$ and hence $D_3$ are positive.
The trace $T_3$  is positive for $x < x_c$ and negative otherwise, with $x_c $ defined by the formula
\begin{equation}
x_c = \lambda-\dfrac{(\lambda-1)(\lambda+d-1)}{ \lambda(2+1/\mu) +(d-1)(1+1/\mu) -1}.
\label{eq:xc}
\end{equation}
We  now  compute the two solutions  $x_\pm$  of the equation $T^2_3=4\Delta_3$. We obtain
\begin{equation}
\label{eq:xpm}
x_\pm = \lambda - \dfrac{\mu}{\rho_3\left(\lambda(2+1/\mu) + (d-1)(1+1/\mu) -1\right) \mp 2 \sqrt \rho_3}.
\end{equation}
Therefore, $P_3$ is an unstable node for $x\le x_-$, an unstable focus for $x \in[x_-;x_c]$ a stable focus for $x \in [x_c;x_+]$ and a stable node for $x \ge x_+$.
As an example,  for the Leith model $d=3, m=11/2$ and $n=1/2$, we obtain $x_- \simeq 1.01$, $x_+ \simeq 2.28$ and $x_c \simeq 1.95$ (and $x_K =5/3 \simeq 1.67$).

\section{Derivation of the autonomous system (\ref{eq:evodim1})--(\ref{eq:evodim2})}
\label{ap:autonomous}
It is straightforward yet slightly tedious to obtain the  system (\ref{eq:evodim1})--(\ref{eq:evodim2}) from the equation (\ref{eq:profiledim}) for the profile function. For thoroughness, the missing steps are provided below.

Equation (\ref{eq:evodim1}) is a simple consequence of the definition of the variables $\rho$,$\sigma$ and $\tau$ as provided by the equations (\ref{eq:changevariable}) and  (\ref{eq:taueta}) ($\rho$ and $\sigma$ are not independent).  Indeed, from the definition (\ref{eq:changevariable}) and the chain rule, we obtain
\begin{equation}
\begin{split}
F^\prime(\eta) &=  \mu \rho^{\mu-1}\rho^\prime(\tau) \dfrac{\d \tau}{\d \log \eta} \eta^{-\lambda-1}-\lambda \rho^\mu\eta^{- \lambda-1}
\\& =\rho^{\mu-1}\eta^{-\lambda-1}\left(\mu \rho^\prime(\tau) - \lambda \rho \right) \sieq\text{on the one hand,}\\
&= \rho^{\mu-1}\eta^{-\lambda-1} \lambda \sigma \sieq \sieq \text{on the other hand.} \\
\end{split}
\end{equation}
For non-vanishing $\rho$, equating the latter two equalities yield Equation (\ref{eq:evodim1}).
Deriving Equation (\ref{eq:evodim2}) is more involving.
First, one gets from Definition (\ref{eq:taueta}) that:
\begin{equation}
 \dfrac{\d\rho }{\d \eta} = \dfrac{\rho^\prime(\tau) }{\rho \eta} \sieq \text{and} \sieq \dfrac{\d \sigma }{\d \eta}  = \dfrac{\sigma^\prime(\tau) }{\rho \eta}.
 \end{equation}
One then needs to plug the definitions (\ref{eq:changevariable}) into Equation (\ref{eq:profiledim}) for the profile function.
The l.h.s of  (\ref{eq:profiledim}) then becomes
\begin{equation}
% \text{l.h.s} =
\dfrac{1}{d+1-m+nx}\left[ x \rho + \lambda \sigma \right]\rho^{\mu-1}\eta^{-\lambda}.
 \label{eq:lhs}
\end{equation}
As for the r.h.s, it now reads in terms of $\lambda$, $\mu$, $n$ and $m$ as follows,
 \begin{equation}
 \begin{split}
% \text{r.h.s} =
\eta^{m-d-(n+1)\lambda-1}&\rho^{(n+1)\mu-2}  \left\lbrace  \left(m-d\right)(1-d) \rho^2  + \lambda^2 \left( n+1-\dfrac{1}{\mu}\right)  \sigma^2  \right. \\
 & \left. + \lambda \left(\left(m-d-\dfrac{\lambda}{\mu}\right) + (1-d)(n+1) \right)\rho\sigma  + \lambda \sigma^\prime(\tau) \right\rbrace.
\end{split}
\label{eq:rhs}
\end{equation}
In order to obtain an autonomous equation for $\sigma^\prime$, one needs to equate  the powers of $\eta$ and $\rho$ that appear on both sides. This prescribes the  following relations  between $(\lambda,\mu)$ and $(m,n)$,
\begin{equation}
 \mu n = 1 \sieq \text{and} \sieq m-d-n\lambda -1 = 0,
\end{equation}
i.e. relations (\ref{eq:lambdamu}).
With such a choice of $\lambda$ and $\mu$, the  l.h.s and r.h.s of the profile equation ---\, as given by the formulae (\ref{eq:lhs}) and (\ref{eq:rhs}) \, --- can be considerably simplified. They now respectively read:
\begin{equation}
\begin{split}
 \text{l.h.s} & =  \dfrac{\mu(x \rho + \lambda \sigma) } {x-\lambda} \sieq \text{and}\\
 \text{r.h.s} & = (1-d) \left(\dfrac{\lambda}{\mu}+1\right)\rho^2 + \lambda^2 \sigma^2  + \lambda\left(2-d+\dfrac{1-d}{\mu}\right)\rho \sigma + \lambda \sigma^\prime(\tau).
\end{split}
\end{equation}
Equating both sides and dividing by $\lambda$   yields Equation  (\ref{eq:evodim2}).

\section{Asymptotics}
\label{ssec:asymp}
We compute $ \Delta x = x_+ - x_-$ in several asymptotics.  When the quantity goes to zero, then $x_\star\to x_K$.
We use the notations of appendix \ref{ap:stability} to compute
\begin{equation}
\begin{split}
 \Delta x = \dfrac{4 \mu}{D_3^{1/2} \left(A^2 D_3 -4 \right) }, &\sieq \text{where} \sieq A = \lambda(2+1/\mu) +(d-1)(1+1/\mu)-1 \\
&\text{and} \sieq D_3 = \dfrac{\mu}{(\lambda-1)(\lambda+d-1)}.
\end{split}
\end{equation}
Recall also that  $\lambda = \mu(x_K-1) + x_K$.

\begin{enumerate}
\item For $x_K \to 1$ (fixed $n$), we obtain  $\lambda \to 1$, $A \to d(1+1/\mu)$, $D_3 \sim \mu/(d (\lambda-1))$. Hence $\Delta x \sim 4\mu^{3/2}(\lambda-1)^{3/2}/\left( d^{1/2} (\mu+1)^2\right) \to 0.$

\item For $x_K \to \infty$ (fixed $n$), we obtain  $\Delta x \sim 4\mu^{1/2} (\mu+1)x_K / \left((2+1/\mu)^2\mu-4\right) \to \infty$.

\item For $n \to \infty$ (fixed $x_K$),  we obtain $\Delta x \sim 4\mu^{3/2} (x_K-1)^{3/2}/(x_K+d-1)^{1/2} \to 0$.
\item  For $n \to 0$ (fixed $x_K$), we obtain $\Delta x \sim \mu^{1/2}(x_K-1) \to \infty$.
\end{enumerate}

For the cases (a) and (c), we therefore obtain that the deviations goes to $0$, \emph{viz.}, $x_\star -x_K \to 0$.
For the two other cases, we cannot conclude. All that we can say is that the deviations cannot grow faster than $x_K$ (case (b)) or $n^{-1/2}$ (case (d)).
\\

\bibliographystyle{iopart-num}
\bibliography{WarmBib}

\providecommand{\newblock}{}
\begin{thebibliography}{10}
\expandafter\ifx\csname url\endcsname\relax
  \def\url#1{{\tt #1}}\fi
\expandafter\ifx\csname urlprefix\endcsname\relax\def\urlprefix{URL }\fi
\providecommand{\eprint}[2][]{\url{#2}}

\bibitem{Leith67}
Leith C~E 1967 {\em Physics of Fluids\/} {\bf 10} 1409--1416
  \urlprefix\url{http://scitation.aip.org/content/aip/journal/pof1/10/7/10.1063/1.1762300}

\bibitem{Kovasznay48}
Kovasznay L~S~G 1948 {\em J. Aeronaut. Sci.\/} {\bf 15} 745--753

\bibitem{hasselman}
Hasselmann S and Hasselmann K 1985 {\em Journal of Physical Oceanography\/}
  {\bf 15} 1369--1377
  \urlprefix\url{http://dx.doi.org/10.1175/1520-0485(1985)015<1369:CAPOTN>2.0.CO;2}

\bibitem{gradual}
L'vov V, Nazarenko S and Rudenko O 2008 {\em Journal of Low Temperature
  Physics\/} {\bf 153} 140--161 ISSN 0022-2291
  \urlprefix\url{http://dx.doi.org/10.1007/s10909-008-9844-0}

\bibitem{lilly1989two}
Lilly D~K 1989 {\em Journal of the Atmospheric Sciences\/} {\bf 46} 2026--2030

\bibitem{nazarenko2007kelvin}
Nazarenko S 2007 {\em JETP letters\/} {\bf 84} 585--587

\bibitem{nazarenko2006KWDAM}
Nazarenko S 2006 {\em Journal of Experimental and Theoretical Physics
  Letters\/} {\bf 83} 198--200 ISSN 0021-3640
  \urlprefix\url{http://dx.doi.org/10.1134/S0021364006050031}

\bibitem{galtier2000weak}
Galtier S, Nazarenko S, Newell A~C and Pouquet A 2000 {\em Journal of Plasma
  Physics\/} {\bf 63} 447--488

\bibitem{Connaughton2004}
Connaughton C and Nazarenko S 2004 {\em Physical review letters\/} {\bf 92}
  044501

\bibitem{bos2012developing}
Bos W~J, Connaughton C and Godeferd F 2012 {\em Physica D: Nonlinear
  Phenomena\/} {\bf 241} 232--236

\bibitem{Grebenev2014}
Grebenev V~N, Nazarenko S~V, Medvedev S~B, Schwab I~V and Chirkunov Y~A 2014
  {\em Journal of Physics A: Mathematical and Theoretical\/} {\bf 47} 025501
  \urlprefix\url{http://stacks.iop.org/1751-8121/47/i=2/a=025501}

\bibitem{sulem1983tracing}
Sulem C, Sulem P~L and Frisch H 1983 {\em Journal of Computational Physics\/}
  {\bf 50} 138--161

\bibitem{BARENBLA_ZELDOVIC}
Barenblatt G and Zeldovich Y {1972} {\em {Annual reviews of fluid mechanics}\/}
  {\bf {4}} {285} ISSN {0066-4189}

\bibitem{Batchelor59}
Batchelor G~K 1959 {\em Journal of Fluid Mechanics\/} {\bf 5}(01) 113--133 ISSN
  1469-7645
  \urlprefix\url{http://journals.cambridge.org/article_S002211205900009X}

\bibitem{Obukhov49}
Obukhov A 1949 {\em Izv. Akad. Nauk. SSSR, Ser. Geogr. and Geophys.\/} {\bf 13}
  58
  \urlprefix\url{http://scitation.aip.org/content/aip/journal/jap/22/4/10.1063/1.1699986}

\bibitem{Corrsin51}
Corrsin S 1951 {\em Journal of Applied Physics\/} {\bf 22} 469--473
  \urlprefix\url{http://scitation.aip.org/content/aip/journal/jap/22/4/10.1063/1.1699986}

\bibitem{Zakharov1967}
Zakharov V~E and Filonenko N~N 1967 {\em Soviet Phys. Dokl.\/} {\bf 11}
  881--883

\bibitem{ZakharovPushkarev1999}
Zakharov V~E and Pushkarev A 1999 {\em Nonl. Proc. Geophys.\/} {\bf 6} 1--10

\bibitem{nazarenko2006sandpile}
Nazarenko S 2006 {\em Journal of Statistical Mechanics: Theory and
  Experiment\/} {\bf 2006} L02002

\bibitem{Pushkarev200098}
Pushkarev A and Zakharov V 2000 {\em Physica D: Nonlinear Phenomena\/} {\bf
  135} 98 -- 116 ISSN 0167-2789
  \urlprefix\url{http://www.sciencedirect.com/science/article/pii/S016727899900069X}

\bibitem{1970zs}
{Zakharov} V~E and {Sagdeev} R~Z 1970 {\em Soviet Physics Doklady\/} {\bf 15}
  439

\bibitem{Iroshnikov}
{Iroshnikov} P~S 1964 {\em Soviet Astronomy\/} {\bf 7} 566

\bibitem{galtier2010nonlinear}
Galtier S and Buchlin {\'E} 2010 {\em The Astrophysical Journal\/} {\bf 722}
  1977

\bibitem{l2010weak}
L'vov V~S and Nazarenko S 2010 {\em Low Temperature Physics\/} {\bf 36}
  785--791

\bibitem{boue2011exact}
Bou{\'e} L, Dasgupta R, Laurie J, L'vov V, Nazarenko S and Procaccia I 2011
  {\em Physical Review B\/} {\bf 84} 064516

\bibitem{FLM:15837}
Nazarenko S and Laval J~P 2000 {\em Journal of Fluid Mechanics\/} {\bf 408}
  301--321 ISSN 1469-7645
  \urlprefix\url{http://journals.cambridge.org/article_S0022112099007922}

\bibitem{Brachet-pancakes}
Brachet M~E, Meneguzzi M, Vincent A, Politano H and Sulem P~L 1992 {\em Physics
  of Fluids A: Fluid Dynamics (1989-1993)\/} {\bf 4} 2845--2854

\bibitem{Cichowlas2005239}
Cichowlas C and Brachet M~E 2005 {\em Fluid Dynamics Research\/} {\bf 36} 239
  -- 248 ISSN 0169-5983 in memoriam: Prof. Richard Bruce Pelz 1957-2002
  \urlprefix\url{http://www.sciencedirect.com/science/article/pii/S0169598305000110}

\end{thebibliography}
\end{document}